
%
%
%
%
%
%
%
%
\magnification =1200
\footline={\hss\tenrm\folio\hss}
\normalbaselineskip = 1.2\normalbaselineskip
\normalbaselines
\parindent=20pt
\overfullrule=0pt

\font\titlefont=cmbx10 scaled \magstep4
\font\authorfont=cmbx10 scaled \magstep2

\def\M{IM}

\def\R{{\Bbb R}}
\def\Z{{\Bbb Z}}
\def\C{{\Bbb C}}
\def\mod{\hbox{ mod }}

\def\O{{\cal O}}
\def\G{{\cal G}}

\def\ref#1{\par\hangindent=1.0cm\hangafter1{\noindent#1}}
\def\tilde{\widetilde}

\def\and{\hbox{\quad and \quad}}

\def\Bbb{\bf}
\catcode`\@=12

%
%

\nopagenumbers

\null
\rightline{DAMTP 94-57}
\rightline{hep-th/9407102}
\vskip 1.5 cm
\centerline{\titlefont Symmetric Monopoles}
\vskip .5 cm
\vskip 1 cm
\centerline{\authorfont N.S. Manton$^1$}
\vskip 1 cm
\centerline{\authorfont M.K. Murray$^2$}
\vskip 1.5cm

\vfill

\centerline{\bf  11 July 1994}

\vfill

\noindent 1. Department of Applied Mathematics and Theoretical Physics,
University of Cambridge, Silver Street, Cambridge CB3 9EW, UK.
{\it nsm10@amtp.cam.ac.uk}

\noindent 2. Department of Pure Mathematics, The University of Adelaide,
Adelaide SA 5005, Australia.  {\it mmurray@maths.adelaide.edu.au}

\vfill\eject
\nopagenumbers

\vskip 1cm

\noindent{\bf Abstract}
We discuss the spectral curves and rational maps associated with $SU(2)$
Bogomolny monopoles of arbitrary charge $k$. We describe the effect on
the rational maps of inverting monopoles in the plane with respect to
which the rational maps are defined, and discuss
the monopoles invariant under such inversion. We define the strongly
centred monopoles, and show they form a geodesic submanifold of the
$k$-monopole moduli space.  The space of strongly centred
$k$-monopoles invariant under the cyclic group of rotations about a
fixed axis, $C_k$, is
shown to consist of several surfaces of
revolution, generalizing the two surfaces obtained by Atiyah and Hitchin
in the 2-monopole case. Geodesics on these surfaces give a novel type
of $k$-monopole scattering.

We present a number of curves in $TP_1$ which
we conjecture are the spectral curves of
monopoles with the symmetries of a regular solid.
These conjectures are based on analogies with Skyrmions.


\vfill\eject

\pageno=1
\footline={\hss\tenrm\folio\hss}

\noindent{\bf 0. Introduction \hfill}

In recent years, there has been considerable interest in Bogomolny
monopoles, which are particle-like solitons in a Yang-Mills-Higgs
theory in three spatial dimensions. In this
paper, we shall only consider $SU(2)$ Bogomolny monopoles in (flat) $\R^3$,
which are the finite energy solutions of the Bogomolny
equations (1.2) [1].  Solutions
are labelled by their magnetic charge, a non-negative integer $k$, and are
physically interpreted as static, non-linear superpositions of $k$
unit charge magnetic monopoles. There is a $4k$-dimensional manifold of
(gauge inequivalent) $k$-monopole solutions, known as the $k$-monopole
moduli space $M_k$, and on this there is a naturally defined
Riemannian metric, which is hyperk\"ahler [2].

For monopoles moving at modest speeds compared with the speed of
light, it is a good approximation to model $k$-monopole dynamics by
geodesic motion on the moduli space $M_k$. This was conjectured some
time ago [3], and the consequences explored in some detail [2, 4]. Very
recently, the validity of the geodesic approximation has been proved
analytically by Stuart [5].

Most studies of Bogomolny monopoles have been concerned either with
the general structure of the $k$-monopole moduli space $M_k$, and its
metric, or with a detailed study of $M_2$ and the geodesics on it,
which describe 2-monopole scattering and bound orbits. Little work has
been done on $k$-monopole dynamics for $k>2$. (The case $k=0$ is
trivial, and if $k=1$, one has a single monopole which moves along a
line at constant speed.) In this paper, we investigate classes of
$k$-monopole solutions which are invariant under various
symmetry groups. We consider monopoles invariant under inversion in a
fixed plane, monopoles invariant under a cyclic group of rotations
about a fixed axis, and monopoles invariant
under the symmetry groups of the regular solids, that is, the tetrahedral,
octahedral and icosahedral groups. The existence of
$k$-monopoles with cyclic
symmetry was previously shown in [6].
A submanifold of the moduli space $M_k$, consisting of all $k$-monopoles
invariant under a fixed symmetry group, is a geodesic submanifold.
We can therefore describe various examples of monopole scattering with
symmetry, by finding geodesics on such submanifolds.

In Sections 1 and 2 we summarize the various ways of characterizing
monopoles, and recall the spectral curves and rational
maps associated with monopoles. In Section 3 we show how the rational
map changes when a monopole is inverted in the plane with respect to
which the rational map is defined, and we investigate the monopoles
which are invariant under this inversion. In Section 4 we consider in
detail the holomorphic geometry associated with the centre of a
monopole. We state (for the first time precisely), both in terms of
the spectral curve and the rational map, the condition for a monopole
to be centred. We also define the total phase of a monopole, and
introduce the notion of a strongly centred monopole -- one whose
centre is at the origin and whose total phase is 1.
In Section 5 we show that the space of
strongly centred $k$-monopoles is a hyperk\"ahler submanifold of
$M_k$, of dimension $4k-4$, which is totally geodesic in $M_k$.

Spectral curves of $k$-monopoles are curves in $TP_1$, the tangent
bundle to the complex projective line, satisfying a number
of constraints. In Section 6 we consider the action of symmetry groups
on general curves in $TP_1$, and present various classes of curves
with cyclic symmetry, and with the symmetries of regular solids.
In Section 7 we consider the
rational maps associated with $k$-monopoles which are symmetric under
the cyclic group $C_k$, and this gives some information on
which curves in $TP_1$ with cyclic symmetry are, in fact, spectral
curves.
The strongly centred monopoles with $C_k$ symmetry are parametrized by
a number of geodesic
surfaces of revolution in the moduli space $M_k$.
We deduce, using the geodesic approximation, a class of novel
$k$-monopole scattering processes, symmetric under the cyclic group
$C_k$. In these, there is a simultaneous collision of $k$ unit charge
monopoles in a
plane, with an $l$-monopole and a $(k-l)$-monopole emerging back-to-back
along the line through the $k$-monopole centre, perpendicular to the
initial plane. Here $l$ can be any integer in the range $0<l<k$. The
outgoing monopole clusters both become axisymmetic about
the line separating them, in the limit of infinite separation.
A purely planar $k$-monopole scattering
process, with $C_k$ symmetry, is also possible.

The investigation of rational maps with cyclic symmetry suggests that
some monopoles and their spectral curves have the symmetry of a
regular solid. We make some precise conjectures about this, and in the
last Section we briefly summarize some results on Skyrmions, another
type of static soliton in a three-dimensional field theory, which
tend to support these conjectures.

Finally a  warning is necessary for the reader who wishes to
delve into the literature on this subject. There are a number of
places in the theory of monopoles where  one has to make choices
and establish conventions.  Most of these are to do with the
orientation of $\R^3$ and the induced complex structure on
the twistor space $T P_1$ of all oriented lines in $\R^3$.
Different authors have made different conventions, and minor
sign inconsistencies can appear to result if the literature is
only read in a cursory manner.

\bigskip
\noindent{\bf  1. Monopoles and the five-fold way \hfill}

As outlined in the Introduction, we wish to present some results on
monopoles and their scattering, with various symmetries.
Before doing this we need to review some of
the theory behind monopoles and
the different points of view from which they can be studied.
Further details of this material can be found in the book
by Atiyah and Hitchin [2] and in the references contained therein.

To define a monopole we start with a pair $(A, \phi)$ consisting of a
connection $1$-form $A$ on ${\Bbb R}^3$ with values in $LSU(2)$, the
Lie algebra of $SU(2)$, and a function $\phi$ (the Higgs field) from
${\Bbb R}^3$ into $LSU(2)$.    The value of the
Yang-Mills-Higgs energy on this pair is defined to be
$$
{\cal E}(A, \phi) = \int_{\Bbb R^3} (|F_A|^2 + |\nabla_A\phi|^2) d^3x
\eqno (1.1)
$$
where $F_A= dA  + A\wedge A$ is the curvature of $A$,
$\nabla_A\phi = d\phi + [A, \phi]$ is the
covariant derivative of the Higgs field, and the norms are taken
using the usual norms on $1$-forms and $2$-forms and an invariant, positive
definite inner product on $LSU(2)$. We call the integrand in ${\cal E}$
the energy density of the monopole.  A standard trick, due to
Bogomolny, can be used to show that the  Yang-Mills-Higgs
energy is minimised by  the solutions of the Bogomolny equations
$$
\star F_A = \nabla_A\phi
\eqno (1.2)
$$
where $\star$ is the usual Hodge star on forms on ${\Bbb R}^3$.
These equations and indeed the energy are invariant under
gauge transformations, where the gauge group
$\G$ of all maps $g$ from ${\Bbb R}^3$   to $SU(2)$ acts by
$$
(A, \phi) \mapsto ( gAg^{-1} - dgg^{-1} \, , \, g\phi g^{-1}).
\eqno (1.3)
$$
Finiteness of the Yang-Mills-Higgs energy, and
the Bogomolny equations, imply certain asymptotic boundary conditions at
infinity  in ${\Bbb R}^3$  on the pair $(A, \phi)$ which are spelt
out in detail in [2]. In particular, $|\phi| \to c$ for some constant
$c$ which cannot change with time. Following [2], we fix $c=1$.

A monopole, then, is a gauge equivalence class of
solutions to the Bogomolny equations subject to these boundary conditions.
In some suitable gauge there is a well-defined Higgs  field
at infinity
$$
\phi^\infty \colon S^2_\infty \to S^2 \subset LSU(2)
\eqno (1.4)
$$
going from the two sphere of all oriented lines through the origin
in $\R^3$ to the unit two-sphere in $LSU(2)$. This Higgs field at infinity
has a degree, or winding number, which, for a solution of
the Bogomolny equations, is a positive integer $k$ called
the magnetic charge of the monopole.

Before discussing the moduli space of all solutions
of the Bogomolny equations we need to be a little more precise
and talk about framed monopoles.  We say a pair $(A, \phi)$ is
framed if
$$
\lim_{x_3 \to \infty} \phi(0, 0, x_3) = \pmatrix{ i & 0 \cr 0 & -i
\cr}.
\eqno (1.5)
$$
The gauge transformations fixing such pairs are those $g$
with  $\lim_{x_3 \to \infty} g(0, 0, x_3) $ diagonal.
Notice that every monopole can be gauge transformed until it is framed.
So the space of monopoles modulo gauge transformations is the same as
the space of framed monopoles modulo
those gauge transformations that fix them.
We define a framed gauge transformation to be one
such that $\lim_{x_3 \to \infty} g(0, 0, x_3) = 1 $.
The quotient of the set of all framed
monopoles of charge $k$ by the group of
framed gauge transformations is a manifold called the moduli space of
(framed) monopoles of charge $k$ and denoted   $M_k$.  The
constant diagonal gauge transformations (a copy of $U(1)$)
act on $M_k$  and the quotient is called the reduced moduli space $N_k$.
This action is not quite free, the element $-1$ acts trivially so
the group $U(1)/\{\pm 1\}$ acts freely on $M_k$.

The dimension of $M_k$ is $4k$ and these
parameters can be understood as follows. In the case that
$k=1$ there is a spherically symmetric monopole called the
Bogomolny-Prasad-Sommerfield (BPS) monopole, or unit charge monopole.
Its Higgs field has a single zero at the origin,
and its energy density is peaked there so it
is reasonable to think of the origin as the
centre or location of the monopole.
The Bogomolny equations are translation invariant so this
monopole can be translated about $\R^3$ and also
rotated by the circle of constant diagonal gauge transformations.
This in fact generates all of $M_1$ which is therefore diffeomorphic to
$\R^3 \times S^1$.   The coordinates on $M_1$ specify
the location of
the monopole and what can be thought of as an
internal phase.

If one was optimistic one would be tempted to think that $M_k$ consists
of unit charge monopoles located at
$k$ points with $k$ internal phases. Even
more optimistically one might hope that, as the Higgs field of the
unit charge monopole vanishes at its point of location,
these $k$ points are  where the  Higgs field vanishes.
This is not true in general but it
is asymptotically correct. There is an asymptotic
region of the moduli space  consisting of approximate superpositions
of $k$ unit charge monopoles located at $k$ widely separated points
and with $k$ arbitrary phases.
Although it is not possible, in general, to assign to a charge $k$ monopole
$k$ points or locations in $\R^3$ it is possible to assign to the
monopole a centre which can be thought of as the average of the locations
of the $k$ particles making up the monopole.   The
important property of this centre is that if we act on the monopole by an
isometry of $\R^3$ the centre moves by
the same isometry.   It is
also possible to assign to a $k$-monopole a total phase; this is
essentially the product of the phases of the $k$ unit charge monopoles.
Whereas in the case of the centre we are essentially adding up
all the individual locations and dividing by $k$, to get a phase
for the monopole we would want to multiply the individual phases and
take a $k$th  root.  Taking a $k$th root of a complex number is, of course,
ambiguous and we have to  content ourselves instead with being able to
define the product of
all the phases -- the total phase.
If we act on the monopole by a constant gauge transformation
corresponding to an element $\mu$ of $U(1)$ then the total phase
changes by $\mu^{2k}$. The power of two here is because it is
$U(1)/\{\pm 1\}$ which acts freely on the monopole, and the power
of $k$ is because this is the total phase.

The natural metric on the moduli space $M_k$ is
obtained from the $L_2$ metric on the fields $(A,\phi )$, taking due
account of gauge invariance. Since a large part of the moduli space $M_k$
describes $k$
well-separated unit charge monopoles, many geodesics on $M_k$
correspond to the scattering of $k$ unit charge monopoles, and we
shall discuss below some particularly symmetric cases of such scattering.

It is not easy to study charge
$k$ monopoles directly in terms of their fields $(A,\phi)$. However, there
are various ways of transforming monopoles to other types of
mathematical objects.
The five approaches to monopoles can be summarised by the diagram:
\smallskip
$$
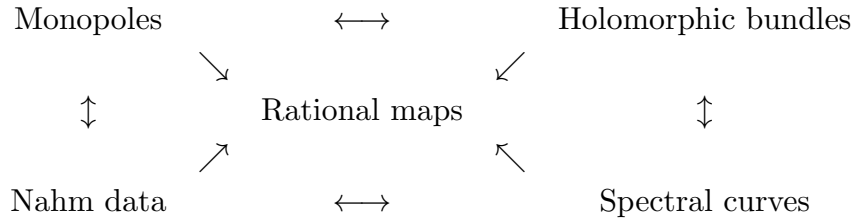
\matrix{
\hbox{Monopoles} &          & \longleftrightarrow  &        &
\hbox{Holomorphic bundles} \cr
                  &\searrow  &                  & \swarrow &
\cr
\updownarrow      &          &\hbox{Rational maps} &        &
\updownarrow         \cr
                  &\nearrow  &                  & \nwarrow &
\cr
\hbox{Nahm data} &          & \longleftrightarrow  &        &
\hbox{Spectral curves} \cr
}
$$
\medskip
In the top left hand corner of the diagram we have monopoles
in $\R^3$ as we have just defined them. There is a twistor  theory
for monopoles and the result of applying this shows that monopoles
are equivalent to a certain class of holomorphic bundles on the so-called
mini-twistor space $TP_1$, the
tangent bundle to the complex projective line $P_1$. This is
indicated by the top horizontal arrow. A careful analysis of the
boundary conditions of the monopole shows that the holomorphic bundle
is determined by an algebraic curve, called the spectral curve.
Monopoles that differ only by a constant diagonal gauge transformation
have the same spectral curve.
These results are due to Hitchin [7, 8].
The vertical arrow on the left is a transformation due to Nahm [9] which
turns the monopole into a solution of an
ordinary differential equation, called Nahm's equation,
on an interval in $\R$. It can be shown that Nahm's equation is equivalent
to a Lax pair equation and hence  one expects to
find  associated to it an algebraic curve and indeed that  curve is
the spectral curve [10].  The relationship between solutions of Nahm's
equations and spectral curves was explained by Hitchin in [8].
Common to all these approaches to monopoles
is a rational map, that is a map
$$
R(z) = {p(z) \over q(z)}
\eqno (1.6)
$$
from $\C$ to $\C \cup \infty$,  where $p$ and $q$ are polynomials.

The holomorphic bundles, spectral curves
or solutions of Nahm's equations that charge $k$ monopoles give rise to all
satisfy some usually rather nasty constraints.
However the rational maps have the enormous advantage that
they are easy to describe. One just writes down a polynomial $p$ of degree
less than $k$ divided by a monic (leading coefficient = 1) polynomial $q$
of degree $k$  which has no factor
in common with $p$. We shall denote by  $R_k$
the space of all these based rational maps.
Donaldson's theorem then assures us that any such rational
map arises from some unique charge $k$ monopole [11].
The disadvantage of this approach is that
there is no explicit way of describing the monopole, given its
rational map.  Moreover defining the rational map,
as we shall see in the next Section,
requires choosing a line and an orthogonal plane in $\R^3$, to
define an isomorphism $\R^3 = \C \times \R$, and this breaks
the symmetries of the problem.  Whereas the Bogomolny
equations are invariant under all the isometries of $\R^3$,
the transformation to a rational map commutes only with
those isometries that preserve the direction of the line.  The action
of these isometries is given as follows.
Let $\lambda \in U(1)$  and  $w \in \C$ define a rotation
and translation, respectively, in the plane $\C$. Let
$t \in \R$ define a translation perpendicular to the plane
and let  $\mu \in U(1)$ define a constant
diagonal gauge transformation.  A rational map $R(z)$ then transforms
under the composition of all these transformations to
$$
\tilde R(z) = \mu^2 \exp(2t) \lambda^{-2k} R(\lambda^{-1}(z-w)).
\eqno(1.7)
$$
Note that this is slightly different to the action described
in [2, eq. (2.11)]. This is one of those places discussed in the
Introduction where different conventions give rise
to different signs.

It would be nice to have a description of the
action of the full isometry group on the space of all rational
maps. In particular, this would settle easily the conjectures
we make below about the existence of monopoles with the symmetries of
regular solids.
We cannot give this. However, we can do two useful things.
Firstly we can describe what happens to the
rational map when we invert a monopole in the chosen plane. Secondly
 we can  determine
the centre and the total phase of a monopole from its rational map.

Our results will be proved below, after we have introduced the
necessary twistor machinery, but they are so simple to
describe  we will do it here.  Let $p(z)/q(z)$ be the
unique representation of the rational map with $q$ monic. Then the
rational map of the inverted monopole is $I(p)(z)/q(z)$, where $I(p)$
is the unique polynomial of degree less than $k$ such that $I(p)(z) p(z)
= 1 \mod q(z)$. If the roots $\beta_1, \dots, \beta_k$ of
$q$ are distinct then $I(p)$ is uniquely determined by the
fact that $I(p)(\beta_i) p(\beta_i) = 1$ for all $i=1,\dots, k$.
For the centre and total phase denote
by $q_0$ the average of all the roots of $q$ and by $\triangle(p,q)$
the resultant of $p$ and $q$.  The  centre of the
monopole is  $(q_0, (1/2k)\log|\triangle(p,q)|)$
and the total phase is $\triangle(p,q) |\triangle(p,q)|^{-1}$.

It is worth noting that these results appear to be inconsistent with
Proposition 3.12 of [2].  There it is argued that if
$$
R(z) = {p(z) \over q(z)} = \sum_i {\alpha_i \over z-\beta_i}
\eqno(1.8)
$$
is the rational map of a charge $k$ monopole, which consists of $k$
well-separated unit charge monopoles, then (using our conventions) the
individual monopoles are approximately located at the points
$(\beta_i, (1/2)\log|{\alpha_i}|)$
and have phases $\alpha_i|\alpha_i|^{-1}$.  This description implies
that inversion fixes the $\beta_i$ and inverts the $\alpha_i$,
whereas we show below in Section 3
that inversion fixes the $\beta_i$ and inverts the $p(\beta_i)$
which are related to the $\alpha_i$ by
$$
\alpha_i = { p(\beta_i) \over \prod_{j\neq i} (\beta_i - \beta_j)}.
\eqno(1.9)
$$
{}From our results it appears more likely that the individual monopoles are
located at the points $(\beta_i, (1/2)\log|p(\beta_i)|)$ and have phases
$p(\beta_i) |p(\beta_i)|^{-1}$ but we have no proof of this.

\bigskip
\noindent{\bf  2. Monopoles and rational maps \hfill}

The rational map of a monopole was originally described by
Donaldson in terms of solutions to Nahm's equations [11].
Hurtubise then showed how it relates to scattering in $\R^3$ and to the
spectral curve of the monopole [12].
It will be convenient for our purposes
to use the  description in terms of spectral curves.

The holomorphic bundle of a $k$-monopole is defined by Hitchin as
follows [7].
Let $\gamma$ be an oriented line in $\R^3$ and let $\nabla_\gamma$
denote covariant differentiation using the connection $A$ along
$\gamma$. Hitchin  considers the  ordinary differential equation
$$
(\nabla_\gamma  - i \phi)v = 0
\eqno(2.1)
$$
where $v : \gamma \to \C^2$.  The vector space $E_\gamma$ of
all solutions to  equation (2.1) is two-dimensional and the union
of all these spaces forms a rank two smooth complex vector bundle $E$
over  the space of all oriented lines in $\R^3$. It is shown by Hitchin
that this space of all oriented lines is a complex manifold,
in fact isomorphic to $ TP_1$.  Hitchin then
shows that $E$ has a holomorphic structure if the monopole
satisfies the Bogomolny equations.  The bundle $E$ has
two holomorphic sub-bundles $E^\pm_1$ defined by defining their fibres
$(E^\pm_1)_\gamma$ at $\gamma$ to be  the space of solutions that decay as
$\pm \infty$ is approached along the line $\gamma$. The set
of $\gamma$ where $(E^+_1)_\gamma = (E^-_1)_\gamma$, so there is a
solution decaying at both ends, forms
a curve $S$ in $TP_1$ called the spectral curve of the monopole.
It is possible to show that a decaying solution decays exponentially
so the spectral curve is also the set of all lines along which there
is an $L_2$ solution. Intuitively
one should think of the spectral lines as being the lines going through
the locations of the monopoles. In the case of charge $1$,
the spectral lines are precisely those going through the centre
of the monopole.

If we describe a typical point in $P_1$ by homogeneous coordinates
$[\zeta_0, \zeta_1]$ then we can cover $P_1$, in the usual way,
by two open sets $U_0$ and $U_1$ where $\zeta_0$ and $\zeta_1$ are
non-zero, respectively. On the set $U_0$ we introduce
the coordinate $\zeta = \zeta_1 / \zeta_0$. Let us also denote
by $U_0$ and $U_1$ the pre-images of these sets under the projection
map from $TP_1$ to $P_1$. Then a tangent vector $\eta {\partial /
\partial \zeta}$ at $\zeta$
in $U_0$ can be given coordinates $(\eta, \zeta)$.
These coordinates allow us to describe an important
holomorphic line bundle $L$ on $TP_1$, introduced by Hitchin [7, pp.587-9],
which has transition function $\exp(\eta/\zeta)$ on the overlap of
$U_0$ and $U_1$. Similarly for
any complex number $\lambda $ we define the bundle $L^\lambda $ by the
transition
function $\exp(\lambda \eta/ \zeta)$.  Finally, if $n$ is any integer
we define the line bundle $L^\lambda (n)$ to be the tensor product
of $L^\lambda $ with the $n$-th power of the pull-back under
projection $TP_1 \to P_1$ of the  dual of the tautological bundle  on $P_1$.
 This has transition function
$\zeta^{-n}\exp(\lambda \eta/\zeta)$. The line bundle $L^0$ is clearly
trivial so we denote it by $\O$, and $L^0(n)$ is denoted by $\O(n)$.

To avoid the potential ambiguity in what we mean by `transition function'
let us be more explicit. The line bundle $L^\lambda (n)$ has non-vanishing
holomorphic sections $\chi_0$ and $\chi_1$ over $U_0$ and
$U_1$ respectively and for points in $U_0 \cap U_1$ these satisfy
$$
\chi_0 = \zeta^{-n} \exp({\lambda \eta\over \zeta}) \chi_1.
\eqno(2.2)
$$
If we consider  an arbitrary holomorphic section $f$ of this line bundle
its restriction to $U_0$ and $U_1$ can be written as
$f = f_0 \chi_0 $ and $f= f_1 \chi_1$ respectively
 where $f_0$ and
$f_1$ are holomorphic functions on $U_0$ and
$U_1$.
As a consequence of equation (2.2) these functions must satisfy
$$
f_0 = \zeta^{n} \exp({-\lambda \eta\over \zeta}) f_1
\eqno(2.3)
$$
at points in the  intersection $U_0 \cap U_1$.
In fact it follows immediately that a holomorphic section of $L^\lambda (n)$
is exactly equivalent to a pair of such holomorphic functions $f_0$
and $f_1$ defined on $U_0$ and $U_1$ and satisfying
(2.3) on $U_0 \cap U_1$.

With these definitions we can present the results of  Hitchin
that we need.  The sub-bundles
$E_1^\pm$ satisfy $E_1^\pm \simeq L^{\pm 1}(-k)$
and the quotients satisfy $E/E_1^\pm \simeq L^{\mp 1}(k)$.
For a framed monopole there are explicit isomorphisms
so we shall write $=$ instead of $\simeq$. The curve $S$
is defined by the vanishing of the map $ E_1^+ \to E/E_1^-$ and hence
by a section of $(E_1^+)^* \otimes E/E_1^- = \O(2k)$.
In terms of the coordinates $(\eta, \zeta)$, $S$ is defined by an
equation of the form
$$
P(\eta, \zeta) \equiv \eta^k + \eta^{k-1} a_1(\zeta)  + \dots + \eta
a_{k-1}(\zeta)
+ a_k(\zeta)= 0,
\eqno(2.4)
$$
where, for $1 \leq r \leq k$, $a_r(\zeta)$ is a polynomial in $\zeta$
of degree at most $2r$.

The space $TP_1$ has a real structure $\tau$, namely, the anti-holomorphic
involution defined by reversing the orientation of the lines in
$\R^3$. In coordinates it takes the form $\tau(\eta, \zeta) =
(-\bar\eta/\bar\zeta^2,
-1/\bar\zeta)$.  The curve $S$ is fixed by this involution, so we say
that it is real. The reality of $S$ implies that for $1 \leq r \leq
k$,
$$
a_r(\zeta) = (-1)^r \zeta^{2r} \overline{a_r(-{1 \over
{\bar\zeta}})}.
\eqno(2.5)
$$
If $k=1$ the spectral curve has the form
$$
\eta = (x_1 + i x_2) -2x_3\zeta - (x_1 - i x_2)\zeta^2
\eqno(2.6)
$$
where $x = (x_1, x_2, x_3)$ is any point in $\R^3$, [7,
eq. (3.2)]. Such a curve is called a
real section as it defines a section of the bundle $TP_1 \to P_1$, and
is real in the sense given above.
In terms of the geometry of $\R^3$ this curve is the set of all
oriented lines through the point $x$, so it is the spectral curve of a BPS
monopole located at $x$. We refer to this curve as the ``star'' at $x$.

In [7, 8] Hitchin lists all the properties that a curve in $TP_1$ has
to satisfy to be a spectral curve.  We are interested in
one of these here. From the definition of the spectral curve
we see that over the spectral curve
the line bundles $E_1^+$ and $E_1^-$ coincide as sub-bundles of $E$; in
particular they must be isomorphic. This is equivalent to
saying that  the line bundle
$E_1^+ \otimes (E_1^-)^* = L^2$ is trivial over the curve
or that it admits a non-vanishing
holomorphic section $s$. The real structure $\tau$ can
be lifted to an anti-holomorphic, conjugate linear
map between the line bundles $L^2$ and $L^{-2}$ and hence
the section $s$ can be conjugated to define a new (holomorphic) section
$\tau(s) = \tau \circ s \circ \tau$ of $L^{-2}$ over
$S$. Tensoring these defines a section
$\tau(s)s$ of $L^{-2} \otimes L^2 = \O$
and because $S$ is compact and connected this is a constant.
Because of the framing this constant will be $1$.
Notice that given only $S$ and the fact that $L^2$ is trivial
over $S$, if we can choose a section $s$ such that $\tau(s)s=1$
 then it is unique up to multiplication by a scalar of
modulus one. This circle ambiguity in the choice of
$s$ corresponds to the framing of the monopole.
In fact, let $\mu$ be a complex number of modulus one corresponding
to a constant diagonal gauge transformation with diagonal
entries $\mu$ and $\mu^{-1}$.
Then it is possible to follow through the proof in Hitchin [7, pp. 593-4]
and show that if we phase rotate a framed  monopole by $\mu$,
the isomorphism $E^+_1 \to L(-k)$ is multiplied by $\mu$
and  the isomorphism
$E^-_1 \to L^*(-k)$
is multiplied  by $\mu^{-1}$. The section $s$ of
$E_1^+ \otimes (E_1^-)^* = L^2$ is therefore  multiplied by
$\mu^2$. Notice that this is consistent with the fact that
the group $U(1)/\{\pm 1\}$  acts freely on the moduli space $M_k$
of framed monopoles.

To define the rational map we fix the fibre $F$ of $TP_1 \to P_1$
where $\zeta = 0$ and identify it with $\C$.
This corresponds to picking an orthogonal splitting of $\R^3$ as
$\C \times \R$. Each point $z$ in $\C$ is identified with
a point in $F$ by setting $z = \eta$,
and hence with an oriented line, the line $\{(x_1,x_2,x_3) \mid x_3
\in \R\}$
with $z = x_1 + ix_2$.
The intersection of $F$ with $S$ defines $k$ points
counted with multiplicity and  $q(z)$ is defined to be the unique monic
polynomial of degree $k$ which has these $k$ points as its roots. Thus
$q(z) = P(z,0)$, where $P$ is given by eq.(2.4).
Recall from eq.(2.3) that a holomorphic
section $s$ of the bundle $L^2$
 is determined locally by functions
$s_0$ and $s_1$, on $U_0 \cap S$ and $U_1 \cap S$ respectively,
such that
$$
s_0(\eta, \zeta) = \exp({-2\eta \over \zeta}) s_1(\eta, \zeta).
\eqno (2.7)
$$
Let $p(z)$ be the unique polynomial of degree $k-1$ such that
$p(z)  = s_0(z , 0) \mod q(z)$.
The rational map of the monopole is then $R(z) = p(z)/q(z)$.
If the roots of $q(z)$ are distinct complex numbers
$\beta_1, \dots , \beta_k$
then the polynomial $p(z)$ is determined by the fact that
$p(\beta_i) = s_0(\beta_i, 0)$ for all $i = 1 , \dots , k$.

Notice that we have departed at this point from the convention
of Hurtubise [12]. There
the rational map is defined using $\tau(s)$. We will see in the
next Section precisely what this means but it may be helpful here to
make a brief remark about the construction of the rational
map as scattering data  in $\R^3$. More details are given
in [12] and  [2].
The points where $S$ intersects $F$ correspond, of course, to the
lines in the $x_3$-direction admitting a solution of eq.(2.1) decaying
at both ends.
Assume these lines are distinct and label them by the corresponding
complex numbers $\beta_i$. Pick for each line a solution $v(\beta_i, x_3)$
decaying at both ends. In the
regions where $x_3$ is large positive and  large negative
there are choices of asymptotically flat gauge such that
$$
 \lim_{x_3 \to \infty} (x_3)^{-k/2} e^{x_3} v(\beta_i, x_3)  =
v_i^+ \pmatrix{1\cr 0\cr}
\eqno(2.8)
$$
and
$$
 \lim_{x_3 \to - \infty} (x_3)^{-k/2} e^{-x_3} v(\beta_i, x_3)  =
 v_i^- \pmatrix{0\cr1\cr}.
\eqno(2.9)
$$
The rational
map with our conventions is determined by
$$
p(\beta_i) = {v^+_i \over v^-_i}.
\eqno(2.10)
$$
This agrees with the results stated in Chapter 16 of ref.[2], although
Hurtubise's conventions give
$p(\beta_i) = v^-_i / v^+_i$.

\bigskip
\noindent{\bf  3. Inverting rational maps}

Consider the inversion  map $I \colon \R^3 \to \R^3$ defined by
$I(x_1, x_2, x_3) = (x_1, x_2, -x_3)$.
This inverts $\R^3$ in the $(x_1, x_2)$ plane.  The inversion map
induces an anti-holomorphic  map on the twistor space $TP_1$  which we
shall denote by the
same symbol and which in the standard coordinates on $TP_1$ is
$$
I(\eta, \zeta) = ({-\bar\eta \over \bar\zeta^2}, {1\over \bar\zeta}).
\eqno (3.1)
$$
To see this note that the real section defined by the
point $I(x_1, x_2, x_3)$ has equation
$$
\eta = (x_1 + i x_2) + 2 x_3 \zeta - (x_1 - ix_2)\zeta^2.
\eqno (3.2)
$$
So a point $I(\eta, \zeta) $ is on this curve
if and only if
$$
{-\bar\eta \over \bar\zeta^2} = (x_1 + ix_2)  + 2 x_3 {1\over \bar\zeta} -
   (x_1 - ix_2){1\over \bar\zeta^2}.
\eqno (3.3)
$$
Conjugating this equation and clearing the denominators we recover
$$
\eta = (x_1 + ix_2)  - 2 x_3 \zeta - (x_1 - ix_2)\zeta^2,
\eqno (3.4)
$$
the equation of the real section defined by the point $(x_1, x_2, x_3)$.
This confirms the formula for $I$.
Notice that $I$ is very similar to the real structure $\tau$; in fact
$I\circ \tau (\eta, \zeta) = (\eta, -\zeta).$

If we invert the monopole defined by the spectral curve $S$ and
section $s$ we obtain
a new curve $I(S)$  and a new section $I(s)$. The definition of
$I(S)$ is straightforward; it is just the image of $S$ under the map
$I$. We shall consider $I(s)$ in a moment. Because $\tau(S) = S$
it follows that $(\eta, \zeta)$ is on $I(S)$ precisely when
$(\eta, -\zeta) $ is on $S$. In particular, the intersection
of $I(S)$ and the fibre $F$ is just the intersection of $S$ and $F$,
since $\zeta = 0$.
So if we denote by $I(p)$ and $I(q)$ the numerator and denominator of
the rational map
for the inverted monopole, we see that $I(q) = q$.

Now consider the section $s$. Notice that both $\tau$ and $I$ interchange
the two coordinate patches $U_0$ and $U_1$. The section $\tau(s)$ is
defined locally by
$$
\tau(s)_0(\eta, \zeta) = \bar{s}_1(\tau(\eta, \zeta))
\quad , \quad   \tau(s)_1(\eta, \zeta) = \bar{s}_0(\tau(\eta, \zeta))
\eqno (3.5)
$$
and the section $I(s)$ by
$$
I(s)_0(\eta, \zeta) = \bar{s}_1(I(\eta, \zeta))
\quad , \quad   I(s)_1(\eta, \zeta) = \bar{s}_0(I(\eta, \zeta)).
\eqno (3.6)
$$
Hence $I(p)$ is defined by
$$
I(p) = I(s)_0(\ , 0) \mod q = \bar{s}_1\circ \tau(\ , 0) \mod q
\eqno (3.7)
$$
using the fact that $\tau(\eta, 0) = I(\eta, 0)$. From
the relation $\tau(s)s = 1$ and equation (3.5) it follows that
$(\bar{s}_1\circ \tau) s_0 = 1$ and hence
$$
\eqalign{
I(p)p & = (\bar{s}_1\circ \tau(\ , 0)) s_0(\ ,0) \mod q \cr
      & = 1 \mod q. \cr}
\eqno (3.8)
$$
Eq.(3.8), and the requirement that the degree of $I(p)$ is less than
$k$, determine $I(p)$ uniquely. If the roots of $q$ are the distinct
complex numbers $\beta_1, \dots, \beta_k$, a useful alternative way
of obtaining $I(p)$ is to notice that it is  the unique
polynomial of degree less than $k$ such
that $I(p)(\beta_i) p(\beta_i) = 1$ for all $i = 1, \dots, k$.

It is interesting to consider the subset of monopoles that are
invariant under inversion. Their spectral curves are given by
polynomials $P(\eta , \zeta)$ which are even in $\zeta$. Their
rational maps satisfy $p^2 = 1$ mod $q$, so that
$I(p) = p$. Let us calculate how many such rational maps there are,
for a given $q$. If the roots $\beta_i$ are
distinct, then $p(\beta_i) = \pm 1$ for all $i$.
To understand what happens when the roots are not necessarily all
distinct, let the distinct roots of $q$ be denoted by
$\beta_1, \dots, \beta_d$ and assume that  $\beta_i$ has multiplicity
$n_i$. Of course, $n_1 + \cdots + n_d = k$.
Now, given a polynomial $p$ of degree less than $k$ we can
associate to it a list of its values and the values of its derivatives
up to order $n_i-1$ at the points $\beta_i$. This defines a linear map
$$
P_k \to \C^k
\eqno(3.9)
$$
from $P_k$, the space of polynomials of degree less than $k$, to $\C^k$.
We claim that this map is a linear isomorphism. Notice that both
these spaces have dimension $k$ so it is enough to check that the map
has no kernel. However if a polynomial $p$ is killed by this map then
it must contain a factor $(z-\beta_i)^{n_i}$ for each $i$. Hence it is
possible to divide $p$ by a polynomial of degree $k$, but this means
$p \equiv 0$. Notice that in the case of distinct roots the construction
of the inverse to the map in equation (3.9) is Lagrange interpolation.
Assume now that  $p^2 = 1 \mod q$. Then  by repeated differentiation
we deduce that $p(\beta_i) = \pm 1$ and that $p^{(1)}(\beta_i), \dots ,
p^{(n_i - 1)}(\beta_i) $ are all zero. Because the map in
equation (3.9) is an isomorphism there is, for each choice of signs
of the $p(\beta_i)$, a  unique  $p$ satisfying these
conditions.
Conversely, given such a  $p$ it follows that $p^2 - 1$ has a zero of
degree at least $n_i$ at $\beta_i$  so that it is divisible
by each of the factors of $q$
and hence  $p^2 = 1 \mod q$. So given $q$ there are $2^d$
possible choices of $p$ making $p/q$
invariant under inversion, where $d$ is
the number of distinct roots of $q$.

Let us denote by $ \M_k$ the set of monopoles
invariant under inversion. In general this
has several components which we
denote by $\{ \M_k^m : m = 0,1,...,k \}$. $\M_k^m $ is the
component of $\M_k$ for
which (while the roots of $q$ are distinct) $m$ values of $p(\beta_i)$
are $+1$, and $k-m$ values are $-1$. Note that $\M_k^m$ and
$\M_k^{k-m}$ are isomorphic; one is obtained from the
other by multiplying $p$ by $-1$. The simplest of the components is
$\M_k^0$. Here $p(z) \equiv 1$, so the rational maps are of the
form
$$
R(z) = {1 \over q(z)}.
\eqno (3.10)
$$
Clearly, $\M_k^0$ is a submanifold of the moduli space $M_k$.
We now prove that
$\M_k$ and hence all of the $\M^m_k$ are submanifolds.
We have seen that the rational maps of the
inversion symmetric monopoles satisfy the equation
$$
 p^2 = 1 \mod q.
\eqno(3.11)
$$
We would like to formulate this equation as the zero set of
a smooth map of maximal rank so that we can
apply the implicit function theorem
to show that $\M_k$ is a submanifold. Note that this equation
holds precisely when $p^2 - 1$ is zero in the $k$-dimensional vector
space
$$
V_q = \C[z] / <q>
\eqno(3.12)
$$
where $<q>$ is the ideal generated by $q$.
This space depends on the point $q$ so we  think of it as the fibre of a
vector bundle $V \to R_k$, where the fibres
in fact depend on $q$ but not on
$p$.  We define a section of $V \to R_k$  by
$$
h(p,q) = p^2 - 1 + <q> \in V_q.
\eqno(3.13)
$$
Now the points of interest, $\M_k$, are where the image of the section $h$
intersects the zero submanifold $V^0 \subset V$, in other words where the
section
vanishes. For a function the condition for its zero set to be a
submanifold is, of course, that the derivative of the function at
every point on the
zero set should be onto. The condition in the case of a section of a
vector bundle is similar but we are interested only in the vertical
component of the derivative.
To define the vertical component of the derivative of
 $h$, note that at a point on the zero
submanifold we can write the tangent space to $V$ as a direct sum of the
tangent space to the fibre,
which is naturally identified with the fibre, and the tangent space to
$V^0$, which is naturally identified with the tangent space to
$R_k$ by the projection $V \to R_k$.  That is
$$
T_{(0, p, q)}V \simeq V_q \oplus T_{(p,q)}R_k.
\eqno(3.14)
$$
Then the tangent map to $h$ at a point $(p,q)$ where
$h(p,q)=0$ can be projected onto $V_q$.
The condition for the zero set of $h$ to be a submanifold is
that this projected map is always onto.

To calculate the tangent map to $h$ note that if
$h(p,q) = 0$ then
$$
p^2 - 1 - \alpha q = 0
\eqno(3.15)
$$
for some polynomial $\alpha$.  The equivalent relation on tangent
vectors is given by differentiating and substituting to obtain
$$
\dot h = 2p \dot p - { (p^2 - 1) \over q} \dot q + <q> .
\eqno(3.16)
$$
Hence the tangent map to $h$ is defined by
$$
T_{(0,p,q)} h \colon ( \dot p , \dot q) \mapsto  ( 2p \dot p - { (p^2 - 1)
\over q}
\dot q + <q>) \oplus (\dot p, \dot q)
\eqno(3.17)
$$
and the composition of this with the projection onto $V_q$ is
$$
(\dot p, \dot q) \mapsto  2p \dot p - { (p^2 - 1) \over q} \dot q + <q>.
\eqno(3.18)
$$
To show that this map is onto it is sufficient to show that
$$
\dot p \mapsto 2 p \dot p + <q>
\eqno(3.19)
$$
is onto, which it is unless we can find a polynomial $\dot p$
of degree less than $k$ such that $ 2 p \dot p = 0 \mod q$. But by unique
factorisation, because $q$ has $k$ factors and $\dot p$ less than $k$,
one of the factors of $q$ would have to occur in $p$. But by assumption
$p$ and $q$ have no common factors so the proof is complete.

Each of the spaces $\M_k^m$ has dimension $2k$, since there are
$2k$ real parameters in $q$ but none in $p$. Moreover, as they
are defined by imposing a symmetry, these spaces are totally geodesic
subspaces of $M_k$.
$\M_k^0$ is naturally diffeomorphic to the moduli space of $k$
flux vortices in the critically coupled abelian Higgs model, since
$k$-vortex solutions are also parametrised by a single monic
polynomial of degree $k$ [13]. However, the metrics in the monopole and
vortex cases will be different.

We are not sure what kind of monopole configurations lie in the
various spaces $\M_k^m$, but we conjecture that for $m = 0$ (or
$m = k$), the energy density is always confined to a finite
neighbourhood of the plane $x_3 = 0$, whereas for $0 < m < k$ it is
possible for there to be monopole clusters arbitrarily far from the
plane $x_3 = 0$, arranged symmetrically with respect to inversion in
this plane. The examples discussed in Section 7 are consistent with
this conjecture. If the roots of $q$ are distinct and well-separated,
then the configurations always consist of a set of unit monopoles with
their centres in the $x_3 = 0$ plane, provided the monopole positions
are as given in the last paragraph of Section 1.

Finally notice that it follows from equations (3.5) and
(3.6) and the fact that $\tau(\eta, 0) = I(\eta, 0)$ that
using $\tau(s)$ to construct the rational map is
the same as using $I(s)$, and hence the $p(\beta_i)$
occuring in the rational map defined using
$\tau(s)$ would be the reciprocal of the $p(\beta_i)$
we use, and would give the rational map as defined by
Hurtubise.

\bigskip
\noindent{\bf 4. Centred monopoles and rational maps \hfill}

We remarked earlier that although the positions and internal phases of the
$k$ \lq particles' in a charge $k$ monopole are only asymptotically
well-defined, every monopole has a well-defined centre and total phase.
This arises naturally in the twistor picture. If $S$ is the spectral curve
of a monopole then it intersects every fibre of $TP_1 \to P_1$ in $k$
points counted with multiplicity. If we add these points together we obtain
a new curve which is given by an equation $\eta + a_1(\zeta) = 0$. This
curve is a real section and hence $a_1$ is of the form
$$
a_1(\zeta) = -k((c_1 + ic_2) - 2c_3\zeta - (c_1 - ic_2)\zeta^2).
\eqno(4.1)
$$
The point $ c= (c_1, c_2, c_3) $ is the centre of the monopole.
To define the total phase requires a little more work.

Let $(\eta_1, \zeta) \dots (\eta_k, \zeta)$ be
the $k$ points in $S$, which are in the fibre of
$TP_1 \to P_1$ over the point $\zeta$. We claim that there
is a well-defined linear map
$$
L^2_{(\eta_1, \zeta)} \otimes \dots \otimes L^2_{(\eta_k, \zeta)} \to
L^2_{(\eta_1 + \dots + \eta_k, \zeta)}
\eqno (4.2)
$$
which, when suitably interpreted, gives rise to a global, holomorphic map.
To define this map we recall from equation (2.7) that a section of $L^2$
is determined locally by a pair of functions
$s_0$ and $s_1$, on $U_0 \cap S$ and $U_1 \cap S$ respectively,
such that
$$
s_0(\eta, \zeta) = \exp({-2\eta \over \zeta}) s_1(\eta, \zeta).
\eqno (4.3)
$$
and we  therefore have
$$
\eqalign{
s_0(\eta_1, \zeta) \dots s_0(\eta_k, \zeta) =&
                \exp({-2(\eta_1 + \dots + \eta_k)\over\zeta})
                s_1(\eta_1, \zeta)\dots s_1(\eta_k, \zeta)    \cr
    = &\exp({2a_1(\zeta)\over \zeta}) s_1(\eta_1, \zeta)\dots s_1(\eta_k,
\zeta).  \cr
}
\eqno (4.4)
$$
 It follows that the functions $s^k_0$
and $s^k_1$ defined by
$$
s^k_0(\eta, \zeta) = s_0(\eta_1, \zeta) \dots s_0(\eta_k, \zeta) \quad{\rm
and}\quad
s^k_1(\eta, \zeta) = s_1(\eta_1, \zeta) \dots s_1(\eta_k, \zeta)
\eqno (4.5)
$$
define a global holomorphic section $s^k$ of $L^2$ over the
real section $\eta + a_1(\zeta) = 0$.  Moreover
because $\tau(s)s = 1$ we must have $\tau(s^k) s^k = 1$.
The bundle $L^2$ over any real section is trivial and
we fix as a choice of trivialisation $f$ over $\eta = k ((c_1 + i c_2) - 2
c_3 \zeta - (c_1 - i c_2)\zeta^2)$
$$
\eqalignno{
f_0(\eta, \zeta) =& \exp 2 k (c_3 + (c_1 - i c_2)\zeta)\cr
f_1(\eta, \zeta) =& \exp 2 k (-c_3 + (c_1 + i c_2)/\zeta).
&(4.6) \cr
}
$$
It is easy to check that this non-vanishing section $f$
satisfies $\tau(f)f = 1$.  If we divide $s^k$ by $f$ we obtain
a holomorphic function which must be constant. In fact because $\tau(s^k)
s^k = 1$
and $\tau(f) f = 1$ this constant is a complex number of modulus $1$.
We define $s^k/f$ to be the total phase of the monopole. Notice that
if we act on the monopole by a constant gauge transformation
$\mu$ then $s$ is replaced by $\mu^2 s$ and the total phase is
multiplied by $\mu^{2k}$.

Some readers may be concerned that our definition of the total phase
depends on the chosen family of trivialisations of $L^2$ over
each real section. It would appear that we could arbitrarily scale
these over each real section and change the definition of the total
phase.  However  it follows from the  construction of $L$ in [7]
 that the group of translations of $\R^3$ acts on  the bundle $L^2$,
covering its action on $T P_1$.  The
family of sections we have described is translation invariant and
therefore unique up to one overall choice of scale.

Let us now see how to construct the centre and total
phase of a monopole from its rational map. Notice first that if
we restrict the equation of the spectral curve to the fibre $\zeta = 0$
we  obtain an equation of the form
$$
\eta^k - k(c_1 + i c_2) \eta^{k-1} + ... = 0
\eqno(4.7)
$$
and hence $c_1 + ic_2$ is the average of the points of
intersection of the spectral curve with $\zeta = 0$ or
the average of the zeros of $q$.

Comparing the construction of the rational map
of a monopole we see that
$$
s^k_0(k(c_1+ic_2), 0) = \prod_i p(\beta_i) = \triangle(p,q)
\eqno(4.8)
$$
the resultant of $p$ and $q$.
It follows that
$$
{s^k\over f} = \triangle(p, q) \exp(-2kc_3).
\eqno(4.9)
$$
So, if $R(z) = p(z)/q(z)$ is the rational map of a monopole with
$q_0$ the average of the roots of $q$ and $\triangle(p,q)$ the resultant
of $p$ and $q$, then the centre of the monopole is
$$
(q_0, (1/2k)\log|\triangle(p,q)|)
\eqno(4.10)
$$
and the total phase is
$$
\triangle(p,q) |\triangle(p,q)|^{-1}.
\eqno(4.11)
$$

It follows that  a monopole is centered if and only if the
zeroes of $q$ sum to zero and $|\triangle(p,q)| = 1$.
It will be useful to use a stronger notion of centring than this.
We call a monopole strongly centred if it is centred and the
total phase is $1$. From what we have just proven a monopole
is strongly centred if and only if its rational map satisfies
$$
q_0 = 0 \quad\hbox{and}\quad \triangle(p, q) = 1.
\eqno(4.12)
$$
We shall denote the  space of strongly centred monopoles by $M_{k,0}$
and show in the  next section that it is a (totally) geodesic submanifold
of the moduli space $M_k$.

\bigskip
\noindent{\bf 5. Strongly centred monopoles \hfill}

Atiyah and Hitchin show that there is a $k$-fold covering of the
$k$-monopole moduli space
$$
\tilde M_k \to M_k,
\eqno(5.1)
$$
and an isometric splitting $\tilde M_k = X \times \R^3 \times S^1$ for some
hyperk\"ahler manifold $X$. We shall construct such a covering and
splitting explicitly using the twistor space of $M_k$. Given this, the
submanifold $X \times \{0\} \times \{1\}$ is clearly a totally geodesic
submanifold of $\tilde M_k$ and because (5.1) is a finite covering the
image of this submanifold  under projection to $M_k$ is also totally
geodesic in $M_k$. We shall show that this is, in fact, the space $M_{k,0}$
of strongly centred monopoles.

Recall from [2] the basic facts about the twistor space of a hyperk\"ahler
manifold. If $M$ is a hyperk\"ahler manifold then the tangent space at any
point of $M$ has complex structures $I, J$ and $K$  defined on it which
satisfy the quaternion algebra relations. In fact we can define a family of
complex structures on the tangent space by forming combinations $aI+bJ+cK$
as long as  $a^2 + b^2 + c^2 = 1$. This  family is clearly a two-sphere.
The union of all these complex structures for all points defines a
two-sphere bundle $Z \to M$ called the twistor space of  $M$. It is in fact
a complex manifold. The details are given on page 39 of Atiyah and
Hitchin's book [2]. They are

\medskip

\noindent{\bf Theorem 1 \quad}  {\sl Let $M$ be a hyperk\"ahler manifold
of real dimension $4n$ and $Z$ its twistor space. Then

\item{(i)} $Z$ is a holomorphic fibre bundle $Z \to  P_1$
over the complex projective line,

\item{(ii)} the bundle admits a family of holomorphic sections each with
normal bundle isomorphic to $\C^{2n} \otimes \O(1)$,

\item{(iii)} there exists a holomorphic section $\omega$ of $\bigwedge^2
T_F^* \otimes
\O(2)$ defining a symplectic form on each fibre $F$,

\item{(iv)} $Z$ has a real structure $\tau$ compatible with (i), (ii) and
(iii) and
covering the antipodal map on $ P_1$.

Conversely, the parameter space of real sections of any complex
manifold $Z$ of complex dimension $2n+1$
satisfying (i) through (iv) is a $4n$-dimensional
manifold  with a natural hyperk\"ahler structure for which $Z$ is
the twistor space. }

\medskip

The importance of this result for us is that the twistor construction of
hyperk\"ahler manifolds  behaves nicely under natural geometric
constructions such as quotients and products. We shall need a number of
instances of this.  The first is that if we have a product $M = M_1 \times
M_2$ of hyperk\"ahler manifolds  then the twistor space $Z$ must be a fibre
product $Z = Z_1 * Z_2$   of the twistor spaces $Z_i \to  P_1 $ of the
$M_i$. Recall that the fibre product is defined by $(Z_1 * Z_2)_{z} =
(Z_1)_{z} \times (Z_2)_{z}$ The converse of this theorem is also true.  If
$Z$ is the twistor space of $M$ and $Z = Z_1 * Z_2$ is a fibre product and
the structures in Theorem 1 decompose  in the natural way then $M$ is a
product of two hyperk\"ahler manifolds $M_1 \times M_2$. The second  is
that if  $\Z_k$, the finite group of $k$-th roots of unity, acts freely on
a twistor space $Z$ commuting with all the  structures in Theorem 1 then it
will act on the corresponding hyperk\"ahler manifold freely and
vice-versa.

The twistor space $Z_k$ of the monopole moduli spaces is constructed on
page 46 of Atiyah and Hitchin's book as follows.

\medskip

\noindent {\bf Theorem 2 \quad} {\sl  The twistor space $Z_k$ of the
moduli space $M_k$
is defined  by taking two copies of $\C \times R_k$ parametrised by
$(\zeta, R(z))$ and $(\tilde\zeta, \tilde R(z))$ and
identifying them  over $\zeta \neq 0$ by
$$
\matrix{
\tilde \zeta = \zeta^{-1} \cr
\tilde q({z \over \zeta^2}) = \zeta^{-2k} q(z) \cr
\tilde p({z\over \zeta^2}) = e^{-2z/\zeta} p(z)\mod q(z). \cr
}
$$
The symplectic form on the fibre is defined by
$$
\omega = \pi \sum {d\tilde\beta_i \wedge {d\tilde p(\tilde \beta_i)} \over
      \tilde p(\tilde \beta_i)} = {\pi\over \zeta^2} \sum {d\beta_i \wedge
           d p( \beta_i) \over  p( \beta_i)}
$$
where the $\tilde \beta_i$ (resp. $\beta_i$) are the roots
of $\tilde q$ (resp. $q$).

The real structure is defined by
$$
\matrix{
\tau(\zeta ) = - \bar\zeta^{-1}\cr
\tau\bigl({p(z)\over q(z)}\bigr) = (-1)^k \bar\zeta^{-2k} \overline{\bigl(
{p(-\bar z\zeta^{-2}) \over   q(-\bar z\zeta^{-2})}\bigr)}.
}
$$
}

\medskip

Define a $k$-fold cover $\tilde R_k$ of $R_k$ by considering
all pairs $(p(z)/q(z), p_0)$ where $p_0$ is a complex number satisfying
$p_0^k = \triangle(p, q)$.  Notice that $\Z_k$ acts freely on
$\tilde R_k$ by multiplying $p_0$ and the quotient is $R_k$.

We can now construct a twistor space $\tilde Z_k$ which is a $k$-fold
cover of $Z_k$ by identifying two copies of $\C \times \tilde R_k$
using the same rules as in Theorem 2 and a rule for identifying
$p_0$ and $\tilde p_0$. To see what that should be note that
$$
\triangle ( \tilde p, \tilde q) = e^{-2kq_0/\zeta} \triangle(p, q).
\eqno(5.5)
$$
A good choice then is to identify $p_0$ and $\tilde p_0$ by
$$
\tilde p_0 = e^{-2q_0/\zeta} p_0.
\eqno(5.6)
$$
Notice that the action of $\Z_k$ extends to a (free) action on $\tilde Z_k$
and that the quotient is $Z_k$. It is straightforward now to lift the
definitions of the symplectic form and the real structure to $\tilde Z_k$
in such a way that
$$
\tilde Z_k \to Z_k
\eqno(5.7)
$$
is a quotient of twistor spaces.  We claim now that
there is a corresponding $k$-fold covering of hyperk\"ahler manifolds
$$
\tilde M_k \to M_k.
\eqno(5.8)
$$
This follows immediately as long as $\tilde M_k$ is non-empty; that
is, as long as we can lift any holomorphic section  $P_1 \to
M_k$ to a holomorphic section $ P_1 \to \tilde M_k$. To see that
this is possible note that the fibration $\tilde Z_k \to Z_k$ restricted to
the image of such a  section is a $k$-fold covering of $ P_1$
without ramification. It must therefore be trivial and hence
the section lifts.
It follows that we  have constructed a $k$-fold covering $\tilde M_k$
of the hyperk\"ahler manifold $M_k$.

If $R_{k,0}$ denotes the strongly centred rational maps we can define an
isomorphism $\tilde R_k \to  R_{k,0} \times \C^\times \times \C$ by
$$
({R(z)}, p_0)  \mapsto (p_0^{-1}R(z + q_0), p_0, q_0)
\eqno(5.9)
$$
where $q_0 = (1/k)\sum \beta_i$ is the average of the roots of $q$.
This map just sends a monopole to the corresponding strongly centred
monopole and the
centre and total phase.  The inverse map is
$$
(R(z),  p_0, q_0) \mapsto (p_0 R(z - q_0), p_0).
\eqno(5.10)
$$
We can also define a subtwistor space $Z_{k,0}\subset Z_k$ by identifying
two copies of $\C \times R_{k,0}$  by the rules of Theorem 2.

Atiyah and Hitchin show that $ \C^\times \times \C$ is
$R_1$ so we can form a twistor space $Z_1$ with two copies of
$\C^\times \times \C \times \C $ and this is the twistor space
of $M_1 = \R^3 \times S^1$.  The map in equation (5.9)
can now be seen to extend to a fibre map
$$
\tilde Z_k \to Z_{k,0} * Z_1.
\eqno(5.11)
$$
It is straightforward to check that this map
is well-defined, i.e. two things identified
in Theorem 2 are still identified after they are mapped by (5.9).
We also need to check that the symplectic form on the fibres of
$\tilde Z_k$ maps to the  product symplectic form on the fibres of
$Z_{k,0} * Z_1$.  To see this note that it is enough
to work on each fibre and show that the pull-back of the
symplectic form on $R_{k,0} \times \C^\times \times \C$
to $\tilde R_k$ under the map
$$
(R(z), p_0)  \mapsto (p_0^{-1}R(z+q_0), p_0, q_0).
\eqno(5.12)
$$
is the symplectic form of $\tilde R_k$.
The pull-back of the symplectic form is
$$
{\pi\over \zeta^2}(\sum
{d(\beta_i - q_0)  \wedge d(p_0^{-1}p(\beta_i))
\over p_0^{-1} p(\beta_i)}
 + k{dq_0 \wedge dp_0 \over p_0}).
\eqno(5.13)
$$
Recall that $p_0^k = \prod p(\beta_i)$ and $q_0 = (1/k)\sum \beta_i$
so that
$$
\sum(\beta_i - q_0 ) = 0 \quad\hbox{and}\quad
k {dp_0 \over p_0} = \sum {dp(\beta_i) \over p(\beta_i)}.
\eqno(5.14)
$$
In the expression (5.13) we need to expand the
term $d(p_0^{-1}p(\beta_i))$. This yields  two terms and in the second
one,  $d(p_0^{-1})p(\beta_i)$, the factor $p(\beta_i)$
cancels with the same factor in the denominator. The sum is then over
$d(\beta_i - q_0)$ which vanishes.  So we have
$$
\eqalignno{
&{\pi\over \zeta^2}
(\sum {d(\beta_i - q_0) \wedge d(p_0^{-1}p(\beta_i))
\over p_0^{-1} p(\beta_i)}
  + k {dq_0 \wedge dp_0 \over p_0}) \cr
&\quad = {\pi\over \zeta^2}(\sum {d\beta_i \wedge dp(\beta_i)  \over
p(\beta_i)} - \sum {dq_0 \wedge dp(\beta_i) \over p(\beta_i)}
+ k { dq_0\wedge dp_0  \over p_0})  \cr
&\quad = {\pi\over \zeta^2}(\sum { d\beta_i\wedge dp(\beta_i) \over
p(\beta_i)}),
&(5.15)}
$$
as required.

We now have an isomorphism of twistor spaces $ \tilde Z_k = Z_{k,0} * Z_1$.
We can define a subset of $ Z_{k,0} * Z_1$ by defining it fibre by
fibre to correspond to the subset  $R_{k,0} \times \{1\} \times \{0\}
\times \C$. It
is clear that this is well-defined. The holomorphic sections which
lie inside this space define the totally
geodesic subspace $M_{k,0} \times \{0\} \times \{1\}$
inside $M_{k,0} \times \C\times \C^\times$ and hence inside $\tilde M_k$.
The image of this under the finite covering $\tilde M_{k,0} $
is also totally geodesic and clearly defines the space of strongly centred
monopoles as it corresponds to monopoles whose rational map is strongly
centered.

\bigskip
\noindent{\bf 6. Symmetric Curves in $TP_1$ \hfill}

In eq.(2.4) we presented the general form of curves in $TP_1$
that occur as spectral curves of charge $k$ monopoles.  The
coefficients $a_r(\zeta)$ must satisfy the reality condition (2.5),
and the curve
is centred at the origin in $\R ^3$ if $a_1(\zeta) = 0$. Here
we shall discuss the form of these curves when they are required to be
invariant under a group of rotations about the origin.

Let us recall that in $TP_1$, the $P_1$ of lines
through the origin are parametrized by $\zeta$ with $\eta = 0$.  The
line in the direction of the Cartesian unit
vector $(x_1, x_2, x_3)$ has
$\zeta = (x_1 + i x_2)/ (1+x_3)$.  It will be important to consider the
homogeneous coordinates $[\zeta_0, \zeta_1]$ on $P_1$, as well as the
inhomogeneous coordinate $\zeta = \zeta_1 /\zeta_0.$

An $SU(2)$ M\"obius transformation on the homogeneous coordinates,
$[\zeta_0, \zeta_1] \rightarrow [\zeta'_0, \zeta'_1]$, of the form
$$
\eqalign{
\zeta'_0 & = - (b + ia) \zeta_1 + (d - ic)\zeta_0 \cr
\zeta'_1 & = (d + ic)\zeta_1 + (b - ia)\zeta_0 \cr
}
\eqno(6.1)
$$
where $a^2 + b^2 + c^2 + d^2 =1$, corresponds to an $SO(3)$
rotation in ${\Bbb R} ^3$.  The rotation is by an angle $\theta$ about
the unit vector $(x_1, x_2, x_3)$, where
$
x_1 \sin {\theta \over 2} = a,\quad x_2 \sin {\theta \over 2} = b,
\quad x_3 \sin {\theta \over 2} = c, \quad
{\cos {\theta \over 2} = d}
$.
The inhomogeneous coordinate $\zeta$ transforms to
$$
{\zeta' = {(d + ic)\zeta + (b - ia) \over -(b + ia)\zeta + (d-ic)}}.
\eqno(6.2)
$$
Since $\eta$ is the coordinate in the tangent space to $P_1$ at $\zeta$,
it follows that if $\zeta$ transforms to $\zeta'$ as in (6.2) then $\eta$
transforms to $\eta'$ via the derivative of (6.2), that is
$$
\eta' = {\eta\over (-(b + ia)\zeta + (d - ic))^2}.
 \eqno (6.3)
$$
A curve $P(\eta,\zeta)= 0$ in $TP_1$ is invariant under the M\"obius
transformation if $P(\eta',\zeta')=0$ is the same curve.  If the curve is
the spectral curve of a monopole, then the monopole is invariant under
the associated rotation.

The simplest group of symmetries is the cyclic group of rotations
about the $x_3$-axis, $C_n$.  The generator is the M\"obius
transformation
$$
\zeta' = e^{2\pi i\over n} \zeta, \quad \eta' = e^{2\pi i\over n}\eta.
\eqno (6.4)
$$
A curve $P(\eta,\zeta) = 0$ is invariant if all terms of $P$ have the same
degree, mod $n$.  A curve of the form (2.4) is $C_n$-invariant if all
terms have degree $k$, mod $n$. In particular, it is $C_k$-invariant
if all terms have degree zero, mod $k$.

For there to be axial symmetry about the $x_3$-axis, with symmetry group
$C_\infty$, the curve must be invariant under $\zeta \rightarrow
e^{i\theta} \zeta, \quad \eta\rightarrow e^{i \theta} \eta$, for all
$\theta$. This requires that all terms in $P(\eta,\zeta)$ have degree $k$.
There is a unique axially symmetric, strongly centred monopole for each
charge $k$. Hitchin has shown that its spectral curve is [7]
$$
\eqalign{
\eta  \prod^{m}_{l=1} (\eta^2 + l^2 \pi^2 \zeta^2) & = 0 \quad \hbox
{for}\quad k=2m +1
\cr
\prod^{m}_{l=0} \left (\eta^2 + (l+ {1 \over 2})^2\pi^2 \zeta^2 \right )
& = 0 \quad \hbox {for}\quad k = 2m + 2. \cr }
\eqno(6.5)
$$
Notice that these curves are not determined by symmetry alone, and that the
coefficients of $P$ are transcendental numbers.  The only curve of the form
(2.4) which has full $SO(3)$ symmetry is $\eta^k = 0$.  This is the
spectral curve of a unit charge monopole at the origin when $k=1$, but for
$k > 1$ it is not the spectral curve of a monopole.

The groups $C_n$ and $C_\infty$ are extended to the dihedral groups $D_n$
and $D_\infty$ by adding a rotation by $\pi$ about the $x_1$-axis.  This
rotation corresponds to the transformation on $TP_1$
$$
\zeta' = {1 \over \zeta}, \quad \eta' = - {\eta \over \zeta^2}.
\eqno(6.6)
$$
Under this transformation, and for any constant $\nu$,
$$
(\eta^2 + \nu \zeta^2)' = {1 \over \zeta^4} (\eta^2 + \nu \zeta^2),
\eqno (6.7)
$$
so each of the axially symmetric monopoles has symmetry group
$D_\infty$.

It is useful to note that by a similar argument to that in
Section 3, the reflection $x_2 \rightarrow -x_2$ corresponds to $\zeta
\rightarrow \overline\zeta , \eta \rightarrow \overline\eta$, so
a curve $P(\eta, \zeta) = 0$ is invariant under this reflection if all
coefficients in $P(\eta, \zeta)$ are real. The axially symmetric
monopoles therefore have this reflection symmetry too.

As an example of finite cyclic or dihedral symmetry, let us consider centred
$k=3$ curves with either $C_3$ or $D_3$ symmetry.  Before imposing the
symmetry, the curves are of the form
$$
\eqalign{
\eta^3 & + \eta(\alpha_4 \zeta^4 + \alpha_3 \zeta^3 + \alpha_2 \zeta^2 +
\alpha_1 \zeta + \alpha_0)
\cr
& + (\beta_6 \zeta^6+ \beta_5 \zeta^5 + \beta_4 \zeta^4 + \beta_3 \zeta^3 +
\beta_2 \zeta^2 + \beta_1 \zeta
+ \beta_0) = 0 \cr}
\eqno(6.8)
$$
subject to the reality conditions
$$
\eqalign{
\alpha_4 & = \overline{\alpha}_0, \ \alpha_3  =-\overline{\alpha}_1, \
\alpha_2 =
\overline{\alpha}_2, \cr
\beta_6 &  = -\overline{\beta}_0,\ \beta_{5} = \overline{\beta}_1, \
\beta_{4} =
-\overline{\beta}_2, \ \beta_3 = \overline{\beta}_3. \cr
}
\eqno (6.9)
$$
$C_3$ symmetry implies that (6.8) reduces to
$$
\eta^3 + \alpha \eta \zeta^2 + \beta \zeta^6 +
\gamma \zeta^3 - \bar{\beta} = 0
\eqno(6.10)
$$
where $\alpha$ and $\gamma$ are real.  By a rotation about
the $x_3$-axis, we can orient the curve so that $\beta$ is real, too,
and then there is reflection symmetry under $x_2 \rightarrow -x_2$.
There is $D_3$ symmetry if $\gamma = 0$; then the curve reduces to
$$
\eta^3 + \alpha \eta \zeta^2 + \beta(\zeta^6 -1) = 0
\eqno(6.11)
$$
with $\alpha$ and $\beta$ real.

The axisymmetric charge $3$ monopole has a spectral curve of type (6.11)
with $\alpha = \pi^2$ and $\beta = 0$.  Also, three
separated unit charge monopoles at the vertices of an equilateral
triangle can have $D_3$ symmetry. The
spectral curve is asymptotic to the product of three stars at
$$
(x_1, x_2, x_3) =\left \{ (a, 0, 0), \quad (a \cos
{2\pi\over 3}, a \sin {2\pi\over 3}, 0),
\quad (a \cos {4\pi \over 3}, a \sin
{4\pi \over 3}, 0) \right
\},
\eqno (6.12)
$$
that is,
$$
(\eta - a (1 - \zeta^2))(\eta - a\omega (1 - \omega \zeta^2)) (\eta
- a\omega^2 (1 - \omega^2 \zeta^2) ) = 0,
\eqno (6.13)
$$
where $\omega = e^{2 \pi i/3}.$  Equation (6.13), when multiplied out, is a
curve of the form (6.11) with $\alpha = 3a^2$ and $\beta =a^3$, or
equivalently $\alpha^3 = 27\beta^2$. We shall find out more about the
spectral curves of charge 3 monopoles with symmetry $C_3$ or $D_3$ when we
consider the rational maps associated with the monopoles (see Section 7).

$C_4$ symmetry is rather a weak constraint on curves with $k = 4$.
$D_4$ symmetry, however, implies that a $k = 4$ curve is of the form
$$
\eta^4 + \alpha \eta^2 \zeta^2 + \beta \zeta^8 + \gamma\zeta^4 + \beta = 0
\eqno (6.14)
$$
with $\alpha,\ \beta$ and $\gamma$ real.
The axisymmetric charge $4$ monopole has this form of spectral curve,
with $\alpha = (5/2) \pi^2, \beta = 0$ and $\gamma = (9/16)
\pi^4$.  Four separated unit charge monopoles at the vertices of
the square $\left \{ (\pm a, 0, 0), (0, \pm a, 0) \right \}$ can have
$D_4$ symmetry. The
spectral curve is asymptotic to a product of stars, and is of the
form (6.14), with $\alpha = 4a^2,\ \beta = - a^4$ and $\gamma = 2a^4$.
After a $\pi/4$ rotation, the monopoles are at $(\pm
a/\sqrt{2}, \pm a/\sqrt{2}, 0)$, and $\alpha = 4a^2, \beta = a^4$ and
$\gamma =2a^4$.

There is another interesting asymptotic monopole configuration, with a
spectral curve of type (6.14).  Consider two well-separated axisymmetric
charge $2$ monopoles, centred at $(0, 0, b)$ and $(0, 0, -b)$, and with
the $x_3$-axis the axis of symmetry.  The spectral curve is asymptotic
to a product of curves associated with the charge $2$ monopoles.
Recall that the spectral curve of a centred axisymmetric charge $2$
monopole is $\eta^2 + \pi^2 \zeta^2 = 0$.  This factorizes as $(\eta + i\pi
\zeta) (\eta - i\pi \zeta) = 0,$ which is a product of stars at the complex
conjugate points $(0,0, \pm i\pi/2).$  Translation by $b$ gives
the curve
$$
\eta^2 + 4b \eta \zeta + (4b^2 + \pi^2)\zeta^2 = 0
\eqno (6.15)
$$
which is the product of stars at $(0,0, b \pm i\pi/2)$.  Similarly,
translation by $-b$ gives
$$
\eta^2 -4b\eta\zeta + (4b^2 + \pi^2) \zeta^2 =0
\eqno (6.16)
$$
and the product of these is the curve
$$
\eta^4 + (2\pi^2 - 8b^2)\eta^2 \zeta^2 + (4b^2 + \pi^2)^2 \zeta^4 = 0.
\eqno (6.17)
$$
Since all terms have degree 4 this curve is axisymmetric; however, the true
spectral curve of the charge $4$ monopole has symmetry $D_4$, as we
shall see in the next Section, becoming axisymmetric only in the limit
of infinite separation.

Let us now investigate the curves in $TP_1$ with the symmetries of a
regular solid. Some of these are special cases of the curves we have
already discussed.  There  are three rotational symmetry groups to
consider, those of a tetrahedron, an octahedron and an icosahedron.
The direct way to construct a symmetric curve is to find M\"obius
transformations which generate the symmetry group, and calculate the
conditions for the curve to be invariant.  For example, a curve of
type (6.14), with $D_4$ symmetry, has octahedral symmetry if it is
invariant under the transformation
$$
\zeta' = {i \zeta + 1 \over \zeta + i} , \quad \eta' =
{-2 \over (\zeta + i)^2 }\eta,
\eqno (6.18)
$$
which corresponds to a $\pi/2$ rotation about the $x_1$-axis, and this
requires that the curve reduces to
$$
\eta^4 + \beta (\zeta^8 + 14 \zeta^4 + 1) = 0.
\eqno (6.19)
$$
A more powerful and less laborious approach is to use the theory of
invariant bilinear forms and polynomials on $P_1$, as expounded in Klein's
famous book [14].

Consider a homogeneous bilinear form $Q_r(\zeta_0, \zeta_1)$ of degree
$r$, and its associated inhomogeneous polynomial $q_r(\zeta)$ defined by
$$
Q_r(\zeta_0, \zeta_1) = \zeta^r_0 q_r (\zeta).
\eqno (6.20)
$$
Generally $q_r$ has degree $r$, but it may have lower degree.  Suppose
$Q_r (\zeta_0, \zeta_1)$ is invariant under a M\"obius transformation of
the form (6.1).  Then $q_r(\zeta)$ transforms in a simple way under the
corresponding transformation (6.2), namely
$$
q'_r (\zeta) = {q_r(\zeta) \over
\left (-(b + ia) \zeta + (d - ic)\right )^r}.
\eqno (6.21)
$$
On the other hand, $\eta$ transforms as in (6.3).  Consider a centred curve
in $TP_1$,
$$
P(\eta, \zeta) \equiv \eta^k + \eta^{k-2} q_4
(\zeta) + \eta^{k-3} q_6 (\zeta) +
\dots +
q_{2k} (\zeta) = 0.
\eqno (6.22)
$$
If, under a M\"obius transformation, each polynomial $q_r(\zeta)$
transforms as in (6.21), and $\eta$ as in
(6.3), then each term in the polynomial $P(\eta, \zeta)$ is multiplied by
the same factor
$\left (-(b + ia) \zeta + (d - ic) \right )^{-2k}$, so the curve is
invariant.  It follows that curves invariant under the rotational
symmetry group
of a regular solid can be constructed from the inhomogeneous
polynomials $q_r$ derived from the bilinear forms $Q_r$ invariant
under the group.

Let $G$ denote the tetrahedral, octahedral or icosahedral group.
Klein has described the ring of bilinear forms, $Inv_G$, which change
only by a constant factor under each
transformation of $G$ $\, - \,$ for each
form these factors define an abelian character of $G$.  Let $Inv_G^\star$
be the subring of strictly invariant forms.  A form $Q$ is in
$Inv_G$ if the roots of the associated polynomial $q$ are
invariant under $G$, that is, if they are the union of a set of
$G$-orbits on $P_1$.

Generic $G$-orbits on $P_1$ consist of $|G|$ points, i.e. 12, 24 and
60 points respectively for the three groups.  The associated forms of
degree $|G|$ are always strictly invariant under $G$, and they span a
vector space of forms, of dimension two.  For each group $G$, there
are also three forms of degree less than $|G|$ associated with special
orbits of $G$, and these generate the ring $Inv_G$.  Let $V$, $E$ and
$F$ be the set of vertices, mid-points of edges, and centres of faces
of the centred regular solid (tetrahedron, octahedron or icosahedron)
invariant under $G$.  Centrally project these points onto the unit
sphere, identified with $P_1$, denoting the resulting sets of points
again by $V, E$ and $F$.  $V$ is a $G$-orbit, so there is a form $Q_V$
in $Inv_G$ and an associated polynomial $q_V$, such that $Q_V$ has
degree $|V|$ and $Q_V = 0$ at all points of $V$.  Similarly, there are
forms and polynomials $Q_E, Q_F$ and $q_E,q_F$.  Table 1 gives the
polynomials $q_V, q_E$ and $q_F$ for the three groups $G$, and a star
indicates that the associated form $(Q_V, Q_E$ or $Q_F)$ is strictly
$G$-invariant.  [A choice of orientation has been made for the solids:
the tetrahedron has its vertices at $(1/\sqrt{3}) (\pm 1, \pm 1, \pm
1)$, with either two or no signs negative; the octahedron has its vertices
on the Cartesian axes; the icosahedron has two vertices on the
$x_3$-axis and is invariant under the dihedral group $D_5$.]

\midinsert

$$\vbox{
\offinterlineskip
\halign{
\strut\vrule \hfil $\;$#$\;$\hfil & \vrule \hfil $\;#\;$
\hfil & \vrule \hfil
$\;#\;$ \hfil & \vrule \hfil $\;#\;$ \hfil \vrule \cr
\noalign{\hrule}
&&& \cr
G & q_{_V} & q_{_E} & q_{_F} \cr
&&& \cr
\noalign{\hrule}
&&& \cr
Tetrahedral & \zeta^4 + 2\sqrt{3} i \zeta^2 + 1& \zeta(\zeta^4 -1)^\star &
\zeta^4 -2\sqrt{3} i \zeta^2 + 1 \cr
&&& \cr
\noalign{\hrule}
&&& \cr
Octahedral & \zeta(\zeta^4 -1) & \zeta^{12} - 33\zeta^8 & \zeta^8 +
14\zeta^4 + 1^\star\cr
&& -33\zeta^4 + 1 & \cr
&&& \cr
\noalign{\hrule}
&&& \cr
Icosahedral & \zeta(\zeta^{10} + 11 \zeta^5 -1)^\star & \zeta^{30} + 522
\zeta^{25}
& \zeta^{20} - 228 \zeta^{15} + 494 \zeta^{10} \cr
& & -10005 \zeta^{20} - 10005 \zeta^{10} &  \cr
&& -522 \zeta^5+ 1^\star & + 228\zeta^5 + 1^\star  \cr
&&& \cr
\noalign{\hrule}
}
}$$

\vskip 5pt
\noindent
{\narrower\smallskip\noindent Polynomials associated with the special
orbits $V, E$ and $F$ of the rotational
symmetry groups of the regular solids.  $A$ star$(\star)$ denotes that
the homogeneous bilinear form $Q$ related to the polynomial $q$ is
strictly invariant.\smallskip}
\medskip
\centerline {{\bf Table 1}}
\endinsert

\bigskip

All the icosahedral forms are strictly invariant because the
icosahedral group $A_5$ is simple, and has no non-trivial abelian
characters.  The tetrahedral forms $Q_V$ and $Q_F$ are not strictly
invariant, but acquire factors of $e^{\pm 2\pi i/3}$under a $2\pi/3$
rotation about a 3-fold symmetry axis; so $Q_V Q_F$ is strictly invariant.
In fact, the polynomial associated with $Q_V Q_F$ is $\zeta^8 + 14\zeta^4
+ 1$, which has octahedral symmetry.  Similarly, the octahedral forms
$Q_V$ and $Q_E$ acquire factors of $-1$ under a rotation by $\pi/2$
around a 4-fold symmetry axis, and $Q_V Q_E$ is strictly invariant.

There are remarkable identities satisfied by the forms  $Q_{V}, Q_{E}$ and
$Q_F$ (which remain true if the forms $Q$ are replaced by the
associated polynomials $q$), namely
$$
\eqalign{
Q_V^3 - Q^3_F - 12\sqrt{3}i \ Q^2_E & = 0 \quad \hbox
{for the tetrahedral group}
\cr
108 \ Q_V^4 -  Q^3_F + Q_E^2 & = 0  \quad \hbox {for the octahedral group} \cr
1728 \ Q^5_V - Q_F^3 - Q^2 _E & = 0 \quad \hbox
{for the icosahedral group}. \cr
}
\eqno (6.23)
$$
These identities occur, because each term is a strictly invariant form
of degree $|G|$, lying in the two-dimensional vector space of forms
associated with the generic $G$-orbits.

We can now write down some examples of invariant curves in $TP_1$,
also satisfying the reality conditions (2.5).
Recall that invariant curves in $TP_1$ must be constructed from
polynomials derived from strictly invariant forms.  The simplest curves
with tetrahedral symmetry are
$$
\eta^3 + i a \zeta(\zeta^4 - 1) = 0
\eqno (6.24)
$$
where $a$ is real.  After a rotation, (6.24) becomes
$$
\eta^3 + a (\zeta^6 + 5\sqrt{2}\zeta^3 - 1) = 0,
\eqno (6.25)
$$
which is of the form (6.10), exhibiting manifest $C_3$ symmetry about the
$x_3$-axis.

The curves in $TP_1$ with $k=4$, and either octahedral
or tetrahedral symmetry, are
$$
\eta^4 + ic\eta\zeta(\zeta^4 - 1) + d(\zeta^8 + 14 \zeta^4 + 1) = 0
\eqno(6.26)
$$
with $c$ and $d$ real.  All such curves have tetrahedral symmetry, and
if $c=0$ the symmetry is octahedral.  Finally, the simplest curves
with icosahedral symmetry are
$$
\eta^6 + a\zeta(\zeta^{10} + 11 \zeta^5 - 1) = 0
\eqno (6.27)
$$
with $a$ real.

We shall discuss in the next Section the possibility that some of
these curves are spectral curves of monopoles.

\bigskip
\noindent{\bf 7. Rational Maps of Symmetric Monopoles, and Monopole
Scattering \hfill}

The advantage of working with the rational maps associated with
monopoles is that there is a $1-1$ correspondence between the maps and
monopoles. Also, cyclic or axial symmetry about the $x_3$-axis, if
present, is manifest.  The information hidden in the rational map is
the full three-dimensional structure of the monopole, and we do not
know which maps, if any, characterise monopoles with the symmetries of
a regular solid.  In this Section, we shall investigate monopoles with
cyclic symmetry, and make some conjectures about monopoles with the
symmetries of regular solids.  We shall also discover some novel types
of geodesic monopole scattering.

Recall that the rational map
of a charge $k$ monopole takes the form
$$
R(z) = {p(z) \over q(z)},
\eqno(7.1)
$$
with $q$ monic of degree $k$ and $p$ of degree less than $k$.
Let $\omega = e^{2\pi i/k}$.
Consider the cyclic group of rotations about the
$x_3$-axis, $C_k$, generated by the transformation $z \rightarrow
z'$, where $z' = \omega z$.
The monopole with rational map $R(z)$ is $C_k$ symmetric if $R(z')$
differs from $R(z)$ only by a constant phase.  We get a class of
charge $k$ monopoles with $C_k$ symmetry for each irreducible
character of $C_k$.
Let us denote the $l$th such class of monopoles by $M_k^l \, (0 \leq
l < k)$.
These are the monopoles whose rational maps are of the form
$$
R(z) = {\mu z^l \over z^k - \nu}
\eqno (7.2)
$$
where $\mu$ and $\nu$ are complex parameters.  For these monopoles,
$R(z') = \omega^l R(z)$.
$\, M_k^l$ is a $4$-dimensional geodesic submanifold of the
moduli space $M_k$, since it arises by imposing a symmetry on
the monopoles.  Its metric is also K\"ahler,
because the set of rational maps
(7.2) is a complex submanifold of the set of all maps (7.1).

Since the strongly centred monopoles are geodesic in the moduli space,
we shall now restrict attention to rational maps of strongly centred,
$C_k$-symmetric monopoles. There is no essential loss of generality in
doing this. For a monopole with a rational map of type
(7.2), the criterion (4.12) for it to be strongly centred reduces to

$$
\mu^{k} \prod^k_{i=1} \left (\beta_i\right)^l  = 1
\eqno (7.3)
$$
where $\{\beta_i: i=1,\dots, k\}$ are the $k$ roots of $z^{k} - \nu =
0$.  Eq.(7.3) is equivalent to $\mu^k \nu^l = \pm 1$, with the lower
sign if both $k$ is even and $l$ odd, and the upper sign otherwise.
The magnitude of $\mu$ is $|\mu| = |\nu|^{-l/k}$, and there are
$k$ choices for the phase of $\mu$.  The rational maps we
obtain are parametrised by several surfaces of revolution. For given
$k$ and $l$ there may be one or more surfaces. For $l=0$, for example,
there are $k$ distinct surfaces, each with $\nu$ a good coordinate;
$\mu$ is a distinct, and constant, $k$th root
of unity on each surface. If $l
\not= 0$, and $k$ and $l$ have highest common factor $h$, there are
$h$ distinct surfaces. As arg $\nu$ increases by $2\pi$, arg $\mu$
decreases by $2\pi l/k$, so arg $\nu$ must increase by $2\pi k/h$ for
$\mu$ to return to its initial value. $\nu$ is therefore a good
coordinate on each surface, but the range of arg $\nu$ is $2\pi k/h$.

For given $k$, and each $l$ in the range $0 \leq l < k$, let us
choose one of the surfaces just described, say, the one containing the
rational map (7.2) with $\nu = 1$ and $\mu = e^{\pi i/k}$ (if $k$ is
even and $l$ odd) or $\mu = 1$ (otherwise). Denote this surface by
$\Sigma_k^l$. If there is
another surface, for a particular value of $l$, then it is isomorphic
to $\Sigma_k^l$, as $\mu$ differs on it simply by a constant phase.
Let us now consider the geodesics on $\Sigma_k^l$, and the associated
$C_k$-symmetric monopole scattering.
The simplest geodesic is when $\nu$ moves along the real axis -- the
monopole then has no angular momentum.

On $\Sigma_k^0$ the rational maps are of the form
$$
R(z) = {1 \over {z^k - \nu}},
\eqno (7.4)
$$
where $\nu$ is an arbitrary complex number. $\Sigma_k^0$ is therefore
a submanifold of the space of inversion symmetric monopoles
$\M_k^0$. For $\nu =0$, the
rational map is that of a strongly centred axisymmetric charge $k$
monopole. If $|\nu|$
is large, there are $k$ well-separated unit charge monopoles at the
vertices of an $k$-gon in ${\Bbb R}^3$, with $x_1+ix_2$ a $k$th root of
$\nu$, and $x_3
= 0$.  The geodesic where $\nu$ moves along the entire real axis
corresponds to a
simultaneous scattering of $k$ unit charge monopoles in the
$(x_1,x_2)$ plane, where the incoming and outgoing trajectories are
related by a $\pi/k$ rotation.  The configuration is instantaneously
axially symmetric when $\nu = 0$.  This kind of symmetric planar
scattering of $k$ solitons has been observed in a number of models,
and can be understood in a rather general way [15].

On $\Sigma_k^l$, with $l \neq 0, \ \nu$ is a
non-zero complex number.  $\nu =
0$ is forbidden, as the numerator and denominator of $R(z)$ would have a
common factor $z^l$.  A simple geodesic is with $\nu$ moving along the
positive real axis, say towards $\nu =0$.  The rational map is
$$
R(z) = {\iota \over \nu^{l/k}} \, {z^l \over {z^k - \nu}}
\eqno (7.5)
$$
where $\iota = e^{\pi i/k}$ (if $k$ is even and $l$ is odd) or $\iota
= 1$ (otherwise).
Then the initial motion is again $k$ unit charge monopoles at the vertices
of a contracting $k$-gon in the $(x_1, x_2)$ plane.  As $\nu
\rightarrow 0$, the map approaches
$$
R(z) = {\iota \over \nu^{l/k}} \, {1 \over z^{k-l}}
\eqno (7.6)
$$
which is the map of a charge $(k-l)$ axisymmetric monopole, centred at
$(0,0, (- l / 2k) \log \nu)$.  This is a positive distance along
the $x_3$-axis as $\nu$ is small.  Following an argument of Atiyah and
Hitchin [2,pp.25-6], we deduce that the charge $k$ monopole has split up,
with one
cluster the charge $k-l$ monopole just described, and a further
cluster (or clusters) near the $x_3$-axis, but not so far up.  In fact,
there is just one other cluster, which is an axisymmetric charge $l$
monopole at a negative distance along the $x_3$-axis.  This is
seen by inverting the original monopole in the $(x_1,x_2)$ plane.  The
procedure described in Section 3 shows that the rational map (7.5)
transforms under inversion to
$$
R(z) = {\tilde\iota \over \nu^{(k-l)/k}} {z^{k-l} \over {z^k - \nu}}
\eqno (7.7)
$$
where $\iota \, \tilde\iota = 1$, because
$$
\eqalign{
\iota \, \tilde\iota \, {z^l \over \nu^{l/k}} \, {z^{k-l} \over
\nu^{(k-l)/k}} & = {z^k \over \nu} \cr
& = 1 \mod z^k-\nu \cr}.
\eqno (7.8)
$$
The inverted monopole therefore has an axisymmetric charge $l$
monopole cluster at $\allowbreak$
$(0,0, - ((k-l)/2k) \log \nu)$, as $\nu \rightarrow 0$, while the
original monopole has this axisymmetric charge $l$
cluster at $\allowbreak$ $(0,0,((k-l)/2k )\log\nu)$.

In the geodesic motion, $k$ unit charge monopoles come in, but the outgoing
configuration is of two approximately axisymmetric monopole clusters,
of charges $k-l$ and $l$, at distances $ld$ and
$-(k-l)d$ along the $x_3$-axis, with $d$
increasing uniformly.  This geodesic motion can, of course, also
be reversed.  The centre of mass of these clusters remains at
the origin.

If $k$ is even and $l=k/2$ then the rational maps, and the
geodesic monopole motion we have described, have an additional
inversion symmetry. $R(z) = z^{k/2}/(\nu^{1/2}(z^k - \nu))$ lies in
the space of inversion symmetric maps $\M_k^{k/2}$, and the
factor $\iota$ makes no essential difference. Consequently, the
outgoing clusters have the same charges and
equal speeds.  Since $\nu$ was assumed to be real, there is reflection
symmetry under
$x_2 \rightarrow -x_2$.  Together with the inversion symmetry, $x_3
\rightarrow -x_3$, we obtain an additional rotational symmetry, by
$\pi$ about the $x_1$-axis.  Hence, monopoles with rational maps of the
form (7.5) have $D_k$ symmetry if $k$ is even and $l = k/2$.
There is also $D_k$ symmetry  if $l=0$, for any $k$.

The surfaces $\Sigma_2^0$ and $\Sigma_2^1$  are the ``rounded cone''
and ``trumpet'' described by Atiyah and Hitchin.  These surfaces are
not isomorphic, but the geodesics with $\nu$ real (on $\Sigma_2^0$)
and $\nu$ real and positive (on $\Sigma_2^1$) \underbar{are}
isomorphic.  Along the first, two unit  charge monopoles scatter
through $\pi/2$ in the $(x_1, x_2)$ plane, and along the second they
scatter through $\pi/2$ in the $(x_1, x_3)$ plane.  There are no
analogous isomorphisms in the higher charge cases.

The general geodesics on the surfaces $\Sigma_k^0$ and
$\Sigma_k^l \ (l \neq 0)$ are presumably analogous to  those on the cone
$\Sigma_2^0$ or trumpet $\Sigma_2^1$.  On $\Sigma_k^0$, they correspond
to $k$ unit charge monopoles scattering in the $(x_1,x_2)$ plane with
net orbital angular momentum.  On $\Sigma_k^l \ (l \neq 0)$, $k$ unit
charge monopoles again come in
with net orbital angular momentum.  If this is small, the geodesic
passes through the trumpet-like surface and two monopole clusters
with magnetic charges $l$ and $k-l$ emerge back-to-back on the
$x_3$-axis.  They also have opposite electric charges, which accounts,
physically, for angular momentum conservation.  If the initial angular
momentum is large, then the geodesic does not pass through the
trumpet, but is reflected, and there are $k$ outgoing unit charge monopoles
in the
$(x_1,x_2)$ plane.

What can we learn about the spectral curves of centred $C_k$-symmetric
monopoles from this discussion of rational maps? First, recall that
monopoles whose rational maps differ only by a phase have the same
spectral curves. We need therefore only consider the chosen surfaces
of rational maps, $\Sigma_k^l$, and their associated monopoles. Let us
also restrict attention to monopoles which are oriented to be
reflection symmetric under $x_2 \rightarrow -x_2$, which requires
$\nu$ to be real, and choose a fixed phase for $\mu$ as $\nu$ varies in
magnitude. This restricts us to $2k-1$ disjoint curves in the
surfaces $\Sigma_k^l$, $(\nu$ real in $\Sigma_k^0, \
 \nu$ positive and $\nu$ negative in $\Sigma_k^l \ (l
\neq 0))$, and these curves are geodesics.
It follows that among the centred curves in
$TP_1$ of the form (2.4) with $C_k$
symmetry and oriented, there are $2k-1$ disjoint loci
of spectral curves. (We refer to a connected,
one-dimensional submanifold of spectral curves as a locus in the space of
curves in $TP_1$.) All these spectral curves will have real
coefficients because of the reflection symmetry. We have been unable
to determine, in general, for which parameter values a curve is a
spectral curve, but we can make some qualitative assertions, based on
knowledge of the asymptotic monopole configurations, and the
axisymmetric configurations.  We restrict our remarks to the cases
$k=3$ and $k=4$.

For $k=3$, and  $l = 0, 1\ \hbox{or}\ 2$, there are five loci of
spectral curves of the form (6.10), with $\beta$ real.   When $l=0$
there is $D_3$-symmetry, so
$\gamma = 0$.  The locus is asymptotic at both ends to $\alpha^3 =
27\beta^2$, with $\beta$ large and positive at one end, and $\beta$
large and negative at the other.  The axisymmetric monopole,  half-way
along the locus, has $\beta=0$ and $\alpha = \pi^2$. Presumably,
$\alpha$ is positive along the whole locus. The four
remaining loci, for $l=1$ and $l=2$, are isomorphic.  This is because
$\nu \rightarrow -\nu$ corresponds to a reflection $x_1 \rightarrow
-x_1$, and because the $l=2$ monopoles are obtained from $l=1$
monopoles by inversion $(x_3 \rightarrow -x_3)$.  Under the first
symmetry $\beta \rightarrow -\beta$, and under the second $\gamma
\rightarrow -\gamma$.  Each of the four loci is asymptotic at one end
to $\alpha^3 = 27\beta^2, \gamma = 0$, with $\beta$ either positive or
negative, and at the other to $\alpha = \pi^2 - 3b^2,\ \beta=0,\ \gamma
= -2b(b^2 + \pi^2)$, with $b$ either positive or negative.  These
latter parameters result from taking the product of the spectral curve
of a unit charge monopole at $(0,0,b)$ with the spectral curve of an
axisymmetric charge 2 monopole at $(0,0, - b/2),$ that is
$$
\eqalign{
P(\eta, \zeta)& = (\eta - 2b\zeta) (\eta^2 + 2b\eta\zeta + (b^2 +
\pi^2) \zeta^2) \cr
& = \eta^3 + (\pi^2 - 3b^2)\eta \zeta^2 - 2b(b^2 +
\pi^2)\zeta^3 = 0 \cr}.
\eqno (7.9)
$$
We note that along these four loci, $\alpha$ passes through$ \ 0 \ $,
and it is possible that the loci pass through four points of the form
$(\alpha, \beta, \gamma) = (0, \pm a, \pm 5\sqrt{2} a)$, for some $a$.
These points correspond to a charge $3$ monopole with tetrahedral
symmetry (and four distinct orientations of the tetrahedron).  From
what we know about Skyrmions and their scattering (see the next
Section), we conjecture that tetrahedral charge 3 monopoles do exist,
and that their spectral curves are on the loci corresponding to
rational maps with $k=3$ and $l=1$ or $2$.

In the case $k=4$, we have seven loci of spectral curves with $C_4$
symmetry.  Only three  of these are essentially different.  The four
corresponding to the rational maps with $l=1$ and $l=3$, and $\nu$
positive or negative, are isomorphic.  The $l=1$ and $l=3$ maps, and
hence the corresponding monopoles and spectral curves, are related by
inversion, and the sign of $\nu$ can be reversed  by a $\pi/4$
rotation.  The spectral curves along these four loci have no higher
symmetry than $C_4$ symmetry.

There are two isomorphic loci of spectral curves corresponding to the
$l=2$ maps.
Here there is inversion symmetry, and the spectral curves are
therefore $D_4$ symmetric and of the form (6.14).  Reversing the sign of
$\nu$ again corresponds to a $\pi/4$ rotation, and $\beta$ changes
sign.  The locus with $\nu$ negative interpolates between the
asymptotic parameter values $\alpha = 4a^2, \beta = a^4, \gamma=2a^4$
with $a$ large (corresponding to four stars at $(1/\sqrt{2}) (\pm
\ a, \pm \ a, 0)$ and the asymptotic values $\alpha = 2\pi^2 - 8b^2,
\ \beta=0, \ \gamma = (4b^2 + \pi^2)^2$ with $b$ large (corresponding to
two axisymmetric charge $2$ monopole clusters on the $x_3$-axis).  Along
the locus,
$\alpha$ changes sign, so it is possible
that when $\alpha=0$ the locus passes through a curve with  $\gamma =
14\beta$.  This spectral curve would correspond to a charge $4$ monopole
with octahedral
symmetry.  Again, by analogy with Skyrmions, we conjecture that such a
monopole does exist.

Finally, there is a single locus corresponding to the $l=0$ maps.
This interpolates between the asymptotic parameter values
$\alpha=4a^2,\ \beta= -a^4,\ \gamma = 2a^4$ and $\alpha = 4a^2, \ \beta
=a^4,\ \gamma = 2a^4$, with $a$ large, and passes through the values
$\alpha = 5\pi^2/2 ,\ \beta=0,\ \gamma = 9 \pi^4 / 16$,
corresponding to  the axisymmetric charge $4$ monopole.  Presumably,
$\alpha$ and $\gamma$  are positive along the entire locus.

In summary, our main result is that in the geodesic scattering of
charge $k$ monopoles, with $C_k$ symmetry and angular momentum zero,
there are two kinds of motion.  First, there is the well-known
possibility of $k$ unit charge monopoles scattering in the $(x_1,
x_2)$ plane through an angle $\pi/k$.  Second, there is the novel
possibility of $k$ unit charge monopoles coming in as before, but
emerging as charge $l$ and charge $k-l$ axisymmetric monopoles moving
back-to-back along the $x_3$-axis.  $l$ can have any integer value in
the range $0 < l < k$.  We conjecture that in the special case $k=3,
\ l=1$ or $2$, the geodesic passes through a configuration with
tetrahedral symmetry, and in the case $k=4,\ l=2$ the geodesic passes
through a configuration with octahedral symmetry.

\bigskip
\noindent{\bf  8. Connections with Skyrmions \hfill}

The Skyrme model is a theory of a scalar field in three spatial
dimensions with values in $S^3$.
Finite energy fields are characterised by their degree $B$,
identified physically with baryon number.  There is a standard
potential energy functional whose minima are the Skyrmions [16].  An
interesting submanifold of Skyrme
fields in $\R^3$ is obtained by calculating
the holonomy of $SU(2)$ Yang-Mills instantons along lines parallel to
the $x_4$-axis in ${\Bbb R}^4$ [17].  Instantons of charge $k$ give rise
to an $(8k-1)$-dimensional manifold of static Skyrme  fields with
baryon number $B = k$, and these Skyrme fields become dynamical if
the instanton moduli are regarded as time-dependent.  Minima of the
potential energy functional, restricted to this submanifold of fields,
give good approximations to the Skyrmions, at least for $B \leq 4$;
the fields have the same symmetries and approximately the same
energies, and are easier to compute [18].

Braaten et al. have found the Skyrmions with $B \leq 6$ [19].
The $B = 1$ Skyrmion (the basic one) is spherically symmetric, like a
unit charge monopole.  The $B =2$ Skyrmion is axially symmetric,
with symmetry group $D_{\infty}$.  This is like the axially symmetric
charge $2$ monopole.  Moreover, two suitably orientated $B = 1$
Skyrmions scatter through $\pi/2$ in a head-on collision, like
monopoles.  This is particularly clear if the Skyrme field is
constrained to the submanifold of instanton-generated fields [17].

The $B=3$ Skyrmion has tetrahedral symmetry.  Three $B = 1$
Skyrmions at the vertices of a large equilateral triangle, and
orientated so that the attractive force between them is  maximised,
relax to the tetrahedral configuration as the triangle becomes
smaller.  Furthermore, we expect that in a head-on collision of three
$B = 1$ Skyrmions, with the same initial configuration, the field
will pass close to the tetrahedron and then emerge as a $B = 1$
Skyrmion and a $B = 2$ Skyrmion moving back-to-back along the
$3$-fold axis of symmetry of the initial triangle.  (For Skyrmions
there is no precise analogue of geodesic scattering at very slow
speeds, so the outgoing Skyrmions would oscillate and emit some
radiation, as do monopoles scattering at finite speed.)  This type of
$3$-Skyrmion scattering has not been simulated numerically, but the
motion we have described seems natural from the point of view of the
instanton-generated $B = 3$ Skyrme fields. It is clearly
analogous to the geodesic scattering of a charge $3$ monopole on the
surface $\Sigma_3^1$, with zero angular momentum.  The existence of
the tetrahedrally symmetric $B = 3$ Skyrmion strengthens the
conjecture that a charge $3$ monopole with this symmetry exists.

The $B=4$ Skyrmion has octahedral symmetry.  Moreover, for
instanton-generated Skyrme fields, there is a scattering channel for
four $B =1$ Skyrmions with tetrahedral symmetry (and maximal
attraction between the Skyrmions) in which they are at the vertices of
a contracting tetrahedron, pass through an octahedrally symmetric
configuration and emerge at the vertices of an expanding tetrahedron
dual to the first [18].  This motion should be a good approximation to the
dynamics of Skyrmions in the full theory.  The monopole analogue of
this motion would be a geodesic scattering of monopoles, where the
spectral curve of the monopoles was at all times of the form (6.26), and
instantaneously had octahedral symmetry.  From the Skyrmion results,
we therefore conjecture that there is a locus of
spectral curves of type (6.26), with $c$ taking all real values and $d$ a
symmetric function of $c$. When $c=0$, the spectral curve and
associated monopole would have octahedral symmetry.

We may also make a conjecture about Skyrmion
scattering, based on our monopole results.  Namely, in a head-on
collision of two $B = 2$ Skyrmions, with their axes of symmetry along
the collision path (and isospin-orientated so that their net pion
dipoles are in opposite directions), we expect the field to pass close
to the octahedral $B=4$ Skyrmion and emerge as four $B = 1$ Skyrmions
at the vertices of an expanding square in a plane perpendicular to
the collision axis.  This motion, or its
reverse, would be analogous to the zero angular momentum geodesic
monopole scattering on $\Sigma_4^2$.

The $B = 5$ and $B = 6$ Skyrmions have rather low symmetry.  In
particular, the $B = 6$ Skyrmion does not have the symmetry of the
icosahedron.  We therefore have no insight from Skyrmions into the
possible existence of icosahedrally symmetric monopoles.

Finally, we remark that the relationship between Skyrmion scattering
and monopole scattering is not systematically understood.  A $B = 1$
Skyrmion has six degrees of freedom, whereas a unit charge monopole
has four.  The moduli space of charge $k$
monopoles has dimension $4k$.  There is a less well-defined moduli
space of Skyrme fields of baryon number $B$, of dimension
$6B$, and a well-defined space of instanton-generated Skyrme
fields of dimension $8B-1$.  It would be interesting if the charge
$B$ monopole moduli space could be identified as a submanifold of
either of these latter spaces.  This is certainly possible for $B
= 2$ [17].

\bigskip
\noindent{\bf  Acknowledgements \hfill}

N.S.M. warmly thanks the Pure Mathematics Department of the University
of Adelaide and Professor Alan Carey for inviting him to visit, and
for financial support.  He also thanks the Physics Department of the
University of Tasmania at Hobart, and Professor Delbourgo, for
hospitality, and the Royal Society and the British Council for travel
grants.

M.K.M thanks the Australian Research Council for support and Michael
Singer for useful conversations.

\bigskip
\noindent{\bf References \hfill }

\item{1.} Bogomol'nyi, E.B.: The stability of classical solutions. Sov. J.
Nucl. Phys. {\bf 24}, 449 (1976)

\item{2.} Atiyah, M.F., Hitchin, N.J.: The geometry and dynamics of
magnetic monopoles. Princeton, NJ: Princeton University Press 1988

\item{3.} Manton, N.S.: A remark on the scattering of BPS monopoles.
Phys. Lett. {\bf 110B}, 54 (1982)

\item{4.} Gibbons, G.W., Manton, N.S.: Classical and quantum dynamics
of BPS monopoles. Nucl. Phys. {\bf B274}, 183 (1986);
\item{} Wojtkowski, M.P.: Bounded geodesics for the Atiyah-Hitchin
metric. Bull. Amer. Math. Soc. {\bf 18}, 179 (1988);
\item{} Bates, L., Montgomery, R.: Closed geodesics on the space of
stable two-monopoles. Commun. Math. Phys. {\bf 118}, 635 (1988);
\item{} Temple-Raston, M., Alexander, D.: Differential cross sections
and escape plots for low-energy solitonic $SU(2)$ BPS magnetic
monopole dynamics. Nucl. Phys. {\bf B397}, 195 (1993)

\item{5.} Stuart, D.: The geodesic approximation for the
Yang-Mills-Higgs equations. U.C. Davis preprint (1994)

\item{6.} O'Raifeartaigh, L., Rouhani, S.: Rings of monopoles with
discrete axial symmetry: explicit solution for $N = 3$. Phys. Lett.
{\bf 112B}, 143 (1982)

\item{7.} Hitchin, N.J.:  Monopoles and geodesics.
Commun. Math. Phys. {\bf 83}, 579 (1982)

\item{8.} Hitchin, N.J.: On the construction of monopoles. Commun.
Math. Phys. {\bf 89}, 145 (1983)

\item{9.} Nahm, W.: The construction of all self-dual monopoles
by the ADHM method. In: Craigie, N.S., Goddard, P. and Nahm, W. (eds.)
Monopoles in Quantum Field Theory. Proceedings,
Monopole meeting, Trieste 1981, pp. 87-94.
Singapore: World Scientific 1982

\item{10.} Hitchin, N.J., Murray, M.K.: Spectral curves and the ADHM method.
Commun. Math. Phys. {\bf 114 }, 463 (1988)

\item{11.} Donaldson, S.K.: Nahm's equations and the classification of
monopoles. Commun. Math. Phys. {\bf 96}, 387 (1984)

\item{12.} Hurtubise, J.: Monopoles and rational maps: a note on a
theorem of Donaldson. Commun. Math. Phys. {\bf 100}, 191 (1985)

\item{13.} Taubes, C.: Arbitrary $N$-vortex solutions to the first order
Ginzburg-Landau equations. Commun. Math. Phys. {\bf 72}, 277 (1980)

\item{14.} Klein, F.: Lectures on the icosahedron. London: Kegan Paul 1913

\item{15.} Kudryavtsev, A., Piette, B., Zakrzewski, W.J.: $\pi/N$
scattering in $2 + 1$ dimensions. Phys. Lett. {\bf 180A}, 119 (1993);
\item{} Dziarmaga, J.: Head-on collision of $n$ vortices. Phys. Rev.
{\bf D49}, 5609 (1994)

\item{16.} Skyrme, T.R.H.: A unified field theory of mesons and
baryons. Nucl. Phys. {\bf 31}, 556 (1962)

\item{17.} Atiyah, M.F., Manton, N.S.: Skyrmions from instantons.
Phys. Lett. {\bf 222B}, 438 (1989); Geometry and kinematics of two
Skyrmions. Commun. Math. Phys. {\bf 152}, 391 (1993)

\item{18.} Leese, R.A., Manton, N.S.: Stable instanton-generated
Skyrme fields with baryon numbers three and four. Nucl. Phys. {\bf A}
(to appear)

\item{19.} Braaten, E., Townsend, S., Carson, L.: Novel structure of
static multisoliton solutions in the Skyrme model. Phys. Lett. {\bf
235B}, 147 (1990)

\bye